\documentclass[aps,reprint]{revtex4-1}
\usepackage{amsmath}
\usepackage{amsfonts}
\usepackage{bm}
\usepackage{braket}
\usepackage{multirow}
\usepackage{algorithm}
\usepackage{algorithmicx}
\usepackage{algpseudocode}
\usepackage{graphicx}
\usepackage{hyperref}

\begin{document}

\title{Linear Response Theory for the Density Matrix Renormalization Group: Efficient Algorithms for Strongly Correlated Excited States}

\author{Naoki Nakatani}
\email{naokin@cat.hokudai.ac.jp}
\affiliation{Catalysis Research Center, Hokkaido University, Kita 21 Nishi 10, Sapporo, Hokkaido 001-0021, Japan}
\author{Sebastian Wouters}
\affiliation{Center for Molecular Modeling, Ghent University, Technologiepark 903, 9052 Zwijnaarde, Belgium}
\author{Dimitri Van Neck}
\affiliation{Center for Molecular Modeling, Ghent University, Technologiepark 903, 9052 Zwijnaarde, Belgium}
\author{Garnet Kin-Lic Chan}
\email{gkchan@princeton.edu}
\affiliation{Department of Chemistry, Princeton University, Frick Chemistry Laboratory, Princeton, NJ 08544, USA}

\begin{abstract}
Linear response theory for the density matrix renormalization group (DMRG-LRT) was first presented in terms of the DMRG renormalization projectors [Dorando et al., J. Chem. Phys. 130, 184111 (2009)]. Later, with an understanding of the manifold structure of the matrix product state (MPS) ansatz, which lies at the basis of the DMRG algorithm, a way was found to construct the linear response space for general choices of the MPS gauge in terms of the tangent space vectors [Haegeman et al., Phys. Rev. Lett. 107, 070601 (2011)]. These two developments led to the formulation of the Tamm-Dancoff and random phase approximations (TDA and RPA) for MPS.
This work describes how these LRTs may be efficiently implemented through minor modifications of the DMRG sweep algorithm, at a computational cost which scales the same as the ground-state DMRG algorithm. In fact, the mixed canonical MPS form implicit to the DMRG sweep is essential for efficient implementation of the RPA, due to the structure of the second-order tangent space.
We present ab initio DMRG-TDA results for excited states of polyenes, the water molecule, and a [2Fe-2S] iron-sulfur cluster.
\end{abstract}

\maketitle

%%
%% Introduction
%%

\section{Introduction}

Entanglement renormalization techniques have recently received much attention as efficient ways to solve the quantum many-body problem in lattice systems, nuclear structure, and quantum chemistry. The density matrix renormalization group (DMRG)\cite{white1992,white1993} was the first of such techniques and is currently the most widely used. Many efficient implementations of the DMRG algorithm exist for ab initio quantum chemistry.\cite{white1999, mitrushenkov2001, chan2002, legeza2003, reiher2007, zgid2008, yanai2009, reiher2010, luo2010, sharma2011, sharma2012, wouters2012} The DMRG algorithm can be understood in terms of its underlying variational ansatz, the matrix product state (MPS),\cite{ostlund1995,rommer1997,cirac2008,schollwock2011} which gives a compact representation of the wavefunction on a one-dimensional lattice graph.

Although the DMRG algorithm has been very successful to investigate ground states of strongly correlated systems,
there are difficulties for excited states. These arise from the fact that the optimal choice of renormalized basis states can differ for the ground and excited states. They can be renormalized separately for each state of interest, yielding an accurate but expensive result.  Conversely, they can be renormalized with a single rotation matrix, which is averaged over the states of interest, yielding a less accurate but cheaper result. This issue is very similar to the one in complete active space self-consistent field (CASSCF) theory,\cite{roos1987} where the optimal choice of single particle orbitals can vary between the ground and excited states.

The first solution is directly targeting the state of interest with a different renormalized basis for each state. This is similar to the state-specific (SS) CASSCF algorithm. As is also the case for the SS-CASSCF algorithm, the renormalized bases of the ground and excited states are no longer orthonormal, and their overlap has to be taken into account.

In the second solution, the so-called state-averaged (SA) algorithm, a common renormalized basis is chosen for both the ground state and all excited states of interest. To compute this renormalized basis, the density matrix is averaged over all targeted states. Since the effective dimension of the renormalized basis per targeted state decreases with each extra targeted state in the SA-DMRG algorithm, it requires a larger number of renormalized basis states to achieve the same accuracy as the ground-state DMRG algorithm. One way to relieve this drawback of the SA-DMRG algorithm is to use the state-averaged harmonic Davidson (SA-HD) DMRG algorithm to target higher excited states directly.\cite{dorando2007}

Recently, excitations have been constructed on top of a reference MPS wavefunction,\cite{dorando2009, kinder2011,haegeman2011, PhysRevB.85.035130,PhysRevB.85.100408,wouters2013,haegeman2013} analogous to the concept of particle-hole excitations on top of a reference Slater determinant. In this post-MPS or post-DMRG theory, the reference MPS wavefunction provides a site-based mean-field ansatz,\cite{B805292C, kinder2011} and excitations consist of ``local'' changes in this mean-field ansatz.\cite{wouters2013} In this way, analogues of the Tamm-Dancoff approximation (TDA),\cite{tamm1945,dancoff1950,wouters2013,PhysRevB.85.035130,PhysRevB.85.100408,haegeman2013} the random phase approximation (RPA),\cite{bohm1953,kinder2011,wouters2013,haegeman2013} and configuration interaction with singles and doubles (CISD)\cite{wouters2013} were derived for an MPS reference wavefunction. The main advantage of the post-DMRG theory is that it allows to derive excited state information from a ground state calculation, without the need to update or augment the renormalized basis states.

For ground-state problems, the DMRG sweep algorithm may be viewed as a particularly robust and efficient way to optimize MPS ground states. For excitations, it is therefore desirable to formulate the DMRG-LRT within a similarly efficient sweep algorithm.
In this work, we start from the equation of motion for a DMRG wavefunction to rederive DMRG-TDA (i.e. CI with singles) and DMRG-RPA within the DMRG language. Subsequently, we provide a step-by-step discussion of the required changes to implement these two methods in an existing DMRG code. Analysis of the computational cost shows that these methods come at the same cost as the DMRG ground-state algorithm. Finally, we compare the performance of SA-DMRG and DMRG-TDA for several excited state problems in ab initio quantum chemistry, to analyze the physical content of the site-based excitations in the DMRG-LRT.

%%
%% Theory Part I
%%

\section{Brief Overview of the DMRG Linear Response Theory}
In this section, we present a brief overview of the DMRG linear response theory (DMRG-LRT), derived from the time-dependent DMRG equation. This was first proposed by Dorando et al.\cite{dorando2009} in the DMRG context, and later recast in the MPS language by Haegeman et al.\cite{PhysRevB.85.035130, PhysRevB.85.100408}, by means of the time-dependent variational principle (TDVP).\cite{haegeman2011} Here, we follow Dorando's work in order to use the DMRG context in what follows.

\subsection{Time-Independent DMRG Equation}
First, we present the time-independent DMRG equation for the usual DMRG algorithm, to introduce definitions and notations. DMRG can be derived from the variational principle for an MPS wavefunction. A generic MPS wavefunction is written as
\begin{equation}
\label{eq:mps_wave_gen}
  \ket{\Psi}
 =\sum_{n_{1} \cdots n_{k}} \mathbf{A}^{n_{1}} \cdots \mathbf{A}^{n_{i}} \cdots \mathbf{A}^{n_{k}} \ket{n_{1} \cdots n_{k}}
\end{equation}
where $n_{i}$ represents the physical index, which has a dimension $d$, and the matrices $\mathbf{A}^{n_{i}}$ are of dimension $M\times M$.
Due to the invariance of the matrix products, the MPS $\mathbf{A}^{n_{i}}$ matrices are not uniquely determined. This extra freedom is called \textit{gauge freedom} in the MPS context. In the DMRG language, the choice of gauge is known as the \textit{canonical form} of the DMRG wavefunction.
%The mixed-canonical form, in order to optimize site $i$, can be written as
In a DMRG sweep the matrices in the MPS wavefunction are optimized one by one. When optimizing the matrix at site $i$, the DMRG expresses the MPS in the mixed canonical form at that site, which can be written as
\begin{equation}
\label{eq:mps_wave}
  \ket{\Psi}
 =\sum_{n_{1} \cdots n_{k}} \mathbf{L}^{n_{1}} \cdots \mathbf{C}^{n_{i}} \cdots \mathbf{R}^{n_{k}} \ket{n_{1} \cdots n_{k}}
\end{equation}
The $\mathbf{L}^{n_{i}}$ and $\mathbf{R}^{n_{i}}$ are now called left- and right-rotation matrices, which satisfy the orthonormality conditions
$\sum_{n_{i}} {\mathbf{L}^{n_{i}}}^{\dag} \mathbf{L}^{n_{i}} = \mathbf{1}$ and
$\sum_{n_{i}} \mathbf{R}^{n_{i}} {\mathbf{R}^{n_{i}}}^{\dag} = \mathbf{1}$, respectively.
The $\mathbf{C}^{n_{i}}$ is the coefficient matrix at site $i$, which is the current target of optimization. The DMRG wavefunction is often written in terms of renormalized states
\begin{equation}
  \ket{\Psi}
 =\sum_{l_{i-1}n_{i}r_{i}} c_{l_{i-1}r_{i}}^{n_{i}} \ket{l_{i-1}n_{i}r_{i}}
\end{equation}
where the left- and right-renormalized bases $\ket{l_{i-1}}$ and $\ket{r_{i}}$ are of the form
\begin{align}
  \ket{l_{i-1}}
&=\sum_{n_{1} \cdots n_{i-1}} \mathbf{L}^{n_{1}} \cdots \mathbf{L}^{n_{i-1}} \ket{n_{1} \cdots n_{i-1}} \\
  \ket{r_{i}}
&=\sum_{n_{i+1} \cdots n_{k}} \mathbf{R}^{n_{i+1}} \cdots \mathbf{R}^{n_{k}} \ket{n_{i+1} \cdots n_{k}}.
\end{align}
Due to the orthonormality of $\mathbf{L}^{n_{i}}$ and $\mathbf{R}^{n_{i}}$, the renormalized bases also satisfy orthonormality conditions: $\braket{l_{i-1}'|l_{i-1}}=\delta_{l_{i-1}'l_{i-1}}$ and $\braket{r_{i}'|r_{i}}=\delta_{r_{i}'r_{i}}$.

In the one-site DMRG algorithm, the Lagrangian $\braket{\Psi|\hat{H}|\Psi}-\lambda(\braket{\Psi|\Psi}-1)$ is minimized for the variations of $\mathbf{C}^{n_{i}}$. Consequently, the DMRG wavefunction is an eigenstate of the effective Schr\"odinger equation
\begin{equation}
\label{eq:dmrg}
%%\mathbf{H}^{[i]}\mathbf{C}^{[i]}=E_{0}\mathbf{S}^{[i]}\mathbf{C}^{[i]}
  \mathbf{H}_{i}\mathbf{C}_{i}=E_{0}\mathbf{S}_{i}\mathbf{C}_{i}
\end{equation}
%where the effective Hamiltonian $\mathbf{H}_{i}$ is spanned by the product basis at site $i$,
%i.e. $\{\braket{l_{i-1}'n_{i}'r_{i}'|\hat{H}|l_{i-1}n_{i}r_{i}}\}$,
%the overlap matrix $\mathbf{S}_{i}$ is an identity matrix if we chose a convenient canonical form,
%and the coefficient tensor $\mathbf{C}_{i}$ is a flattened view of $\{\mathbf{C}^{n_{i}}\}$.
where the effective Hamiltonian $\mathbf{H}_{i}$ and the overlap matrix $\mathbf{S}_{i}$ are spanned by the product basis at site $i$,
i.e. $\{\braket{l_{i-1}'n_{i}'r_{i}'|\hat{H}|l_{i-1}n_{i}r_{i}}\}$ and $\{\braket{l_{i-1}'n_{i}'r_{i}'|l_{i-1}n_{i}r_{i}}\}$, respectively,
and the coefficient tensor $\mathbf{C}_{i}$ is a flattened view of $\{\mathbf{C}^{n_{i}}\}$.
%An important feature of the DMRG algorithm is that, because of the use of the mixed-canonical form, the overlap matrix $\mathbf{S}_{i}$ always the identity matrix. This leads to good conditioning and is one of the reasons for the high efficiency of the DMRG algorithm. An important step of the sweep algorithm is therefore to transform the mixed-canonical form from one site to the next. In DMRG language, this is equivalent to the procedure of decimation, which ensures that we keep only $M$ renormalized states both in the left-block $\{\ket{l_{i-1}}\}$ and the right-block $\{\ket{r_{i}}\}$ so that the wavefunction optimization can be performed with a polynomial complexity $\mathcal{O}(M^{3}k^{3})$. Once $\mathbf{C}_{i}$ is determined from Eq.~\eqref{eq:dmrg}, the left- or right-rotation matrices can be found by means of a QR decomposition or singular value decomposition (SVD), such that
An important feature of the DMRG algorithm is that, because it uses the mixed-canonical form, the overlap matrix $\mathbf{S}_{i}$ always the identity matrix. This leads to good numerical conditioning and is one of the reasons for the high efficiency of the DMRG algorithm. An important step of the sweep algorithm is therefore to transform the mixed-canonical form from one site to the next. In DMRG language, this is equivalent to the procedure of decimation, which ensures that we keep only $M$ renormalized states both in the left-block $\{\ket{l_{i-1}}\}$ and the right-block $\{\ket{r_{i}}\}$ so that the wavefunction optimization can be performed with a polynomial complexity $\mathcal{O}(M^{3}k^{3})$. Once $\mathbf{C}_{i}$ is determined from Eq.~\eqref{eq:dmrg}, the left- or right-rotation matrices can be found by means of a QR decomposition or singular value decomposition (SVD), such that
\begin{equation}
  \label{gauge-shift}
  \mathbf{C}_{i}=\mathbf{L}_{i} \bm{\Lambda}_{i}=\bm{\Lambda}_{i-1} {\mathbf{R}_{i}}
\end{equation}
where the left- and right rotation matrices $\mathbf{L}_{i}$ and $\mathbf{R}_{i}$ are rectangular matrices having dimensions of $dM \times M$ and $M \times dM$, respectively, and $\bm{\Lambda}_{i}$ is an $M \times M$ matrix.
%Reshaping $\mathbf{L}_{i}$ to $\mathbf{L}^{n_{i}}$ and $\mathbf{R}_{i}$ to $\mathbf{R}^{n_{i}}$, we can then replace $\mathbf{C}^{n_{i}}$ by the new left (or right) rotation matrix, thereby transferring the mixed-canonical form from one site to the left (or right).
Reshaping $\mathbf{L}_{i}$ to $\mathbf{L}^{n_{i}}$ and $\mathbf{R}_{i}$ to $\mathbf{R}^{n_{i}}$, we can then replace $\mathbf{C}^{n_{i}}$ by the new left (or right) rotation matrix, thereby transferring the mixed-canonical form one site to the left (or right).

\subsection{Time-Dependent DMRG Equation and Linear Response Theory (LRT)}
Now, we briefly introduce the time-dependent DMRG equation to derive the DMRG-LRT.\cite{dorando2009, haegeman2011, kinder2011, wouters2013} The time-dependent DMRG wavefunction can be written similar to Eq.~\eqref{eq:mps_wave},
\begin{equation}
  \ket{\Psi(t)}
 =\sum_{n_{1} \cdots n_{k}} \mathbf{L}^{n_{1}}(t) \cdots \mathbf{C}^{n_{i}}(t) \cdots \mathbf{R}^{n_{k}}(t) \ket{n_{1} \cdots n_{k}}
\end{equation}
Minimizing the Dirac-Frenkel action $\braket{\Psi|i\partial/\partial t - \hat{H}|\Psi}$ for the variations of $\mathbf{C}^{n_{i}}(t)$, gives an equation of motion for the DMRG wavefunction at site $i$
\begin{equation}
\label{eq:tddmrg}
  i\mathbf{S}_{i}(t)\frac{\partial}{\partial t} \mathbf{C}_{i}(t)
 =\mathbf{H}_{i}(t) \mathbf{C}_{i}(t)
\end{equation}
where $\mathbf{S}_{i}(t)$ and $\mathbf{H}_{i}(t)$ are the overlap and the effective Hamiltonian spanned by the product states $\ket{l_{i-1}(t)n_{i}r_{i}(t)}$ at site $i$.
It should be noted that $\mathbf{H}_{i}(t)$ is always time-dependent, even if the full Hamiltonian operator $\hat{H}$ is time-independent, because the renormalized states $\ket{l_{i-1}(t)}$ and $\ket{r_{i}(t)}$ are time-dependent.

%First, we would like to briefly introduce the DMRG linear response theory (DMRG-LRT) to figure out the relation to the Random Phase Approximation (RPA) for MPS.\cite{kinder2011,wouters2013,bohm1953} The detailed derivation of the DMRG-LRT was given in the previous work.\cite{dorando2009}
We can now introduce DMRG-LRT. To investigate the linear response from the time-dependent perturbation
\begin{equation}
\label{eq:pt_ext}
  \hat{V}_{\text{ext}}(t) = \hat{V}e^{-i \omega t} + \hat{V}^{*}e^{i \omega t},
\end{equation}
we introduce a perturbation expansion for the DMRG wavefunction:
\begin{equation}
\label{eq:mps_1st}
\begin{split}
  \mathbf{L}_{i}(t)
&=( \mathbf{L}_{i}^{(0)}+\lambda\mathbf{L}_{i}^{(1)}(t)+\cdots ) e^{-iE_{0}t} \\
  \mathbf{C}_{i}(t)
&=( \mathbf{C}_{i}^{(0)}+\lambda\mathbf{C}_{i}^{(1)}(t)+\cdots ) e^{-iE_{0}t} \\
  \mathbf{R}_{i}(t)
&=( \mathbf{R}_{i}^{(0)}+\lambda\mathbf{R}_{i}^{(1)}(t)+\cdots ) e^{-iE_{0}t} \\
\end{split}
\end{equation}
where the zeroth order elements $\mathbf{L}_{i}^{(0)}$, $\mathbf{C}_{i}^{(0)}$, and $\mathbf{R}_{i}^{(0)}$ are determined by the time-independent DMRG equation, Eq.~\eqref{eq:dmrg}, and are from now on fixed for each site $i$. We next insert Eq.~\eqref{eq:mps_1st} in Eq.~\eqref{eq:tddmrg}, and collect the first order terms per Fourier mode. This gives a pair of frequency-dependent DMRG linear response equations for each site~$i$
\begin{equation}
\label{eq:realresp}
  (\mathbf{H}_{i}^{(0)} - (E_{0} - \omega)\mathbf{1}) \mathbf{C}_{i,+\omega}^{(1)}
 =-\mathbf{Q}_{i}^{C} (\Delta\mathbf{H}_{i,+\omega}^{(1)} + \mathbf{V}_{i}^{(1)}) \mathbf{C}_{i}^{(0)}
\end{equation}
\begin{equation}
\label{eq:imagresp}
  (\mathbf{H}_{i}^{(0)} - (E_{0} + \omega)\mathbf{1}) \mathbf{C}_{i,-\omega}^{(1)}
 =-\mathbf{Q}_{i}^{C} (\Delta\mathbf{H}_{i,-\omega}^{(1)} + \mathbf{V}_{i}^{(1)*}) \mathbf{C}_{i}^{(0)}
\end{equation}
where $\mathbf{Q}_{i}^{C} = \mathbf{1} - \mathbf{C}_{i}^{(0)} {\mathbf{C}_{i}^{(0)}}^{\dag}$ and $\Delta\mathbf{H}_{i,\pm\omega}^{(1)}$
is the first order change of the effective Hamiltonian at site~$i$.

The first order DMRG wavefunction is of the explicit form
\begin{equation}
\label{eq:psi1st}
\begin{split}
  \ket{\Psi_{\pm\omega}^{(1)}}
=&\sum_{n_{1} \cdots n_{k}}[ \mathbf{L}_{\pm\omega}^{n_{1}(1)} \cdots \mathbf{C}^{n_{i}(0)} \cdots \mathbf{R}^{n_{k}(0)} + \cdots \\
 &                         + \mathbf{L}^{n_{1}(0)} \cdots \mathbf{C}_{\pm\omega}^{n_{i}(1)} \cdots \mathbf{R}^{n_{k}(0)} + \cdots \\
 &                         + \mathbf{L}^{n_{1}(0)} \cdots \mathbf{C}^{n_{i}(0)} \cdots \mathbf{R}_{\pm\omega}^{n_{k}(1)} ] \ket{n_{1} \cdots n_{k}}
\end{split}
\end{equation}
where the $+\omega$ and $-\omega$ components correspond to the forward and backward propagating parts in time, respectively.
%In previous work,\cite{dorando2009, haegeman2011, wouters2013} it was argumented that no parametrization freedom is lost by constraining
%It is well known that in linear response theory, no loss of freedom obtains from using intermediate normalization, where $\ket{\Psi^{(0)}}\perp\ket{\Psi^{(1)}}$. The projector $\mathbf{Q}_{i}^{C}$ in Eqs.~\eqref{eq:realresp}, \eqref{eq:imagresp}, ensures that ${\mathbf{C}_{i,\pm\omega}^{(1)}}^{\dag}\cdot\mathbf{C}_{i}^{(0)}=0$, however further conditions must also be placed on the first order changes in the left- and right-rotation matrices. In previous work,\cite{dorando2009}, it was shown that these conditions are
It is well known that in linear response theory, no loss of variational freedom occurs by restricting $\ket{\Psi^{(0)}}\perp\ket{\Psi^{(1)}}$. The projector $\mathbf{Q}_{i}^{C}$ in Eqs.~\eqref{eq:realresp}, \eqref{eq:imagresp}, ensures that ${\mathbf{C}_{i,\pm\omega}^{(1)}}^{\dag}\cdot\mathbf{C}_{i}^{(0)}=0$. In addition, further conditions must also be placed on the first order changes in the left- and right-rotation matrices. In previous work,\cite{dorando2009}, it was shown that these conditions are
\begin{equation}
\sum_{n_{k}} {\mathbf{L}^{n_{k}(1)}}^{\dag} \mathbf{L}^{n_{k}(0)} = \bm{0} \label{left-orthoSeb}
\end{equation}
to the left of the current site (site $i$ in Eq.~\eqref{eq:psi1st}) and
\begin{equation}
\sum_{n_{k}} \mathbf{R}^{n_{k}(0)} {\mathbf{R}^{n_{k}(1)}}^{\dag} = \bm{0} \label{right-orthoSeb}
\end{equation}
to the right of this site.
%Together with requiring that $\ket{\Psi^{(0)}} \perp \ket{\Psi^{(1)}_{\pm \omega}}$,
As is further shown in Refs. \onlinecite{haegeman2011} and \onlinecite{wouters2013},
all gauge freedom in the first order wavefunction is fixed this way.

%The efficient solution of the LR equations \eqref{eq:realresp}, \eqref{eq:imagresp} is achieved by solving them at each site in a DMRG sweep. As discussed in the previous section, this requires shifting the mixed canonical form from one site to the next. In Ref. \onlinecite{dorando2009}, this was achieved by writing down the linear response equations for the left and right rotation matrices in terms of the first order density matrices, such as $\sum_{n_{i}} \mathbf{C}_{\pm\omega}^{n_{i}(1)} {\mathbf{C}^{n_{i}(0)}}^{\dag} +h.c.$. Here, we formulate the transformation in terms of the coefficient matrices directly.
An efficient solution of the LR equations \eqref{eq:realresp}, \eqref{eq:imagresp} is achieved by solving them at each site in a DMRG sweep. As discussed in the previous section, this requires shifting the mixed canonical form from one site to the next. In Ref. \onlinecite{dorando2009}, this was achieved by writing down the linear response equations for the left and right rotation matrices in terms of the first order density matrices.
For the left rotation matrix, the response equation is of the form
\begin{equation}
\label{eq:leftresp}
   \left( \mathbf{D}_{L}^{(0)} - \bm{\sigma}_{p}\mathbf{1} \right) \mathbf{l}_{p}^{(1)} = - \mathbf{Q}_{i}^{L} \mathbf{D}_{L}^{(1)} \mathbf{l}_{p}^{(0)}
\end{equation}
where the zeroth and the first order density matrices, $\mathbf{D}_{L}^{(0)}$ and $\mathbf{D}_{L}^{(1)}$ are given by
\begin{equation}
\begin{split}
   \mathbf{D}_{L}^{(0)} &= \text{Tr}_{R} \left[ \mathbf{C}_{i}^{(0)}{\mathbf{C}_{i}^{(0)}}^{\dag} \right] \\
   \mathbf{D}_{L}^{(1)} &= \text{Tr}_{R} \left[ \mathbf{C}_{i}^{(1)}{\mathbf{C}_{i}^{(0)}}^{\dag} \right] + h.c.
\end{split}
\end{equation}
$\mathbf{l}_{p}^{(0)}$ and $\mathbf{l}_{p}^{(1)}$ are the $p$-th renormalized basis states in the zeroth and the first order spaces, respectively,
and the projector $\mathbf{Q}_{i}^{L}$ is defined as
\begin{equation}
   \mathbf{Q}_{i}^{L} = \mathbf{1} - \sum_{p}^{M} \mathbf{l}_{p}^{(0)}{\mathbf{l}_{p}^{(0)}}^{\dag} = \mathbf{1} - \mathbf{L}_{i}^{(0)}{\mathbf{L}_{i}^{(0)}}^{\dag}.
\end{equation}
To solve \eqref{eq:leftresp}, we get the first order left rotation matrix as $\mathbf{L}_{i}^{(1)} = \{ \mathbf{l}_{p}^{(1)} \}_{1 \leq p \leq M}$.

Here, we formulate the transformation in terms of the coefficient matrices directly.
In the zeroth order wavefunction, the mixed-canonical form is shifted between sites with Eq.~\eqref{gauge-shift}:
\begin{equation}
  \mathbf{C}_{i}^{(0)}\mathbf{R}_{i+1}^{(0)}
 =\mathbf{L}_{i}^{(0)}\bm{\Lambda}_{i}\mathbf{R}_{i+1}^{(0)}
 =\mathbf{L}_{i}^{(0)}\mathbf{C}_{i+1}^{(0)}
\end{equation}
where the left factors of each term are $dM \times M$ matrices, the right factors $M \times dM$ matrices, and where $\bm{\Lambda}_{i}$ is the $M \times M$ matrix describing the choice of the gauge at the $i$-th boundary:
\begin{equation}
  \bm{\Lambda}_{i}
 ={\mathbf{L}_{i}^{(0)}}^{\dag} \mathbf{C}_{i}^{(0)}
 =\mathbf{C}_{i+1}^{(0)} {\mathbf{R}_{i+1}^{(0)}}^{\dag}.
\end{equation}

%The mixed-canonical form can also be shifted in the first order part of the wavefunction. Consider
To shift the mixed canonical form in the first order part of the wavefunction, consider
all first order contributions on the relevant sites. The transformation we would like to achieve is
\begin{equation}
\label{eq:gauge1}
  \mathbf{C}_{i}^{(1)} \mathbf{R}_{i+1}^{(0)} + \mathbf{C}_{i}^{(0)} \mathbf{R}_{i+1}^{(1)} 
 =\mathbf{L}_{i}^{(1)} \mathbf{C}_{i+1}^{(0)} + \mathbf{L}_{i}^{(0)} \mathbf{C}_{i+1}^{(1)}.
\end{equation}
Given the left (right) part, can we find a solution for the right (left) part of the equation? We start with the right-to-left case. Multiplying Eq. \eqref{eq:gauge1} with ${\mathbf{R}_{i+1}^{(0)}}^{\dag}$ to the right gives:
\begin{equation}
\label{eq:gauge_bwd}
  \mathbf{C}_{i}^{(1)}
 =\mathbf{L}_{i}^{(1)} \mathbf{C}_{i+1}^{(0)} {\mathbf{R}_{i+1}^{(0)}}^{\dag} + \mathbf{L}_{i}^{(0)} \mathbf{C}_{i+1}^{(1)} {\mathbf{R}_{i+1}^{(0)}}^{\dag}.
\end{equation}
Multiplying Eq. \eqref{eq:gauge1} with $\mathbf{Q}_{i+1}^{R} = \mathbf{1} - {\mathbf{R}_{i+1}^{(0)}}^{\dag} \mathbf{R}_{i+1}^{(0)}$ to the right and $\bm{\Lambda}_{i}^{-1} {\mathbf{L}_{i}^{(0)}}^{\dag}$ to the left gives:
\begin{equation}
\label{eq:right1st}
  \mathbf{R}_{i+1}^{(1)}
 =\bm{\Lambda}_{i}^{-1} \mathbf{C}_{i+1}^{(1)} \mathbf{Q}_{i+1}^{R}
\end{equation}
The first order terms on the LHS of Eq. \eqref{eq:gauge1} can hence be calculated if the RHS is known. Analogously, the left-to-right case yields the equations:
\begin{eqnarray}
\mathbf{C}_{i+1}^{(1)} & = & {\mathbf{L}_{i}^{(0)}}^{\dag} \mathbf{C}_{i}^{(1)} \mathbf{R}_{i+1}^{(0)} + {\mathbf{L}_{i}^{(0)}}^{\dag} \mathbf{C}_{i}^{(0)} \mathbf{R}_{i+1}^{(1)} \label{eq:gauge_fwd}\\
  \mathbf{L}_{i}^{(1)} & = & \mathbf{Q}_{i}^{L} \mathbf{C}_{i}^{(1)} \bm{\Lambda}_{i}^{-1} \label{eq:left1st}
\end{eqnarray}
with $\mathbf{Q}_{i}^{L} = \mathbf{1} - \mathbf{L}_{i}^{(0)} {\mathbf{L}_{i}^{(0)}}^{\dag}$. These relations enable to change the canonical form of the first order wavefunction to perform an efficient sweep algorithm for DMRG-LRT. Note that Eqs. \eqref{eq:gauge_bwd} and \eqref{eq:gauge_fwd} are very similar to the guess wavefunction transformation of the one-site DMRG algorithm.\cite{chan2002} The left and right projectors $\mathbf{Q}_{i}^{L}$ and $\mathbf{Q}_{i}^{R}$ hold the first order rotation matrices to be orthogonal to the zeroth order contributions, so that they satisfy Eqs. \eqref{left-orthoSeb} and \eqref{right-orthoSeb}, respectively.

\subsection{Tamm-Dancoff Approximation (TDA) and Random Phase Approximation (RPA)}
%The random phase approximation (RPA) for DMRG was previously derived in the MPS context as the linearization of the time-dependent variational principle.\cite{kinder2011, wouters2013, haegeman2013} The Tamm-Dancoff approximation (TDA) can be understood as a variational approximation of RPA. In this work, we will present a sweep algorithm for RPA and TDA in the DMRG context. In order to do so, we first repeat the RPA equations in the MPS context:
The LR equations define a first order wavefunction which determines the dynamic response of observables, such as spectral functions. From the poles of the response, we obtain excited states and their eigenvalues. Formulating the determination of the poles as an eigenvalue problem yields the Tamm-Dancoff and Random Phase Approximations to excited states in DMRG. An explicit route to derive the DMRG-TDA and DMRG-RPA eigenvalue equations is to use a linearization of the time-dependent variational principle,\cite{kinder2011, wouters2013, haegeman2013} from which the TDA can be understood as a variational approximation to RPA.  Our objective here is to formulate an efficient sweep algorithm to solve the DMRG-TDA and DMRG-RPA equations. To do so, we first recall the DMRG-RPA eigenvalue problem:
\begin{equation}
\label{eq:dmrgrpa}
\begin{pmatrix}
\mathbf{H}     & \mathbf{W}     \\
\mathbf{W}^{*} & \mathbf{H}^{*} \\
\end{pmatrix}
\begin{pmatrix}
\mathbf{X} \\
\mathbf{Y} \\
\end{pmatrix}
=
\omega
\begin{pmatrix}
\mathbf{S} &  \mathbf{0}     \\
\mathbf{0} & -\mathbf{S}^{*} \\
\end{pmatrix}
\begin{pmatrix}
\mathbf{X} \\
\mathbf{Y} \\
\end{pmatrix}
\end{equation}
where $\mathbf{H}$, $\mathbf{W}$, and $\mathbf{S}$ are $dM^{2}k \times dM^{2}k$ matrices. Their $dM^{2} \times dM^{2}$ block components (depending on the site indices $i$ and $j$) are
\begin{equation}
  \mathbf{H}_{ij}
 =\braket{\partial_{i}\Psi^{(0)}|\hat{H} - E_{0}|\partial_{j}\Psi^{(0)}}
\end{equation}
\begin{equation}
  \mathbf{W}_{ij}
 =\braket{\partial_{i}\partial_{j}\Psi^{(0)}|\hat{H} - E_{0}|\Psi^{(0)}}
\end{equation}
\begin{equation}
  \mathbf{S}_{ij}
 =\braket{\partial_{i}\Psi^{(0)}|\partial_{j}\Psi^{(0)}}
\end{equation}
where $\ket{\partial_{i}\Psi^{(0)}}$ and $\ket{\partial_{i}\partial_{j}\Psi^{(0)}}$ are the first and second order derivatives of the DMRG wavefunction with respect to the site components $\mathbf{A}_{i}$
\begin{align}
  \ket{\partial_{i}\Psi^{(0)}}
&=\hat{Q}_{i}^{(1)}\frac{\partial}{\partial \mathbf{A}_{i}} \ket{\Psi^{(0)}} \\
  \ket{\partial_{i}\partial_{j}\Psi^{(0)}}
&=\hat{Q}_{ij}^{(2)}\frac{\partial^{2}}{\partial \mathbf{A}_{i} \partial \mathbf{A}_{j}} \ket{\Psi^{(0)}}.
\end{align}
The site components $\mathbf{A}_{i}$ can be in one of the canonical forms $\{\mathbf{L}_{i}^{(0)}, \mathbf{C}_{i}^{(0)}, \mathbf{R}_{i}^{(0)}\}$, depending on the canonical form of $\ket{\Psi^{(0)}}$.
The operator $\hat{Q}^{(n)}$ projects the bare derivative to the $n$-th order subspace orthogonal to all lower order subspaces, as discussed later.
The $dM^{2}k$ dimensional vectors $\mathbf{X}=\{\bm{x}_{1},\cdots,\bm{x}_{k}\}$ and $\mathbf{Y}=\{\bm{y}_{1},\cdots,\bm{y}_{k}\}$ in Eq. \eqref{eq:dmrgrpa} represent the forward and backward propagating RPA amplitudes.
%of the wavefunction, respectively, in Eq.~(\ref{eq:psi1st}).

If $\ket{\Psi^{(0)}}$ is accurate enough, the $\mathbf{W}$ matrix vanishes because $(\hat{H} - \hat{E}_{0})\ket{\Psi^{(0)}} \approx 0$, which leads to the Tamm-Dancoff approximation (TDA)
\begin{equation}
  \mathbf{H} \mathbf{X} = \omega \mathbf{S} \mathbf{X}
\end{equation}
which is a variational method which targets excitations in the subspace spanned by $\left\{ \ket{\partial_{i}\Psi^{(0)}} \right\}$.

\subsection{Non-Redundant Parameterizations of the First and Second Order Spaces}
The space spanned by the bare derivatives contains lower order derivatives. The first order space $\left\{ \partial/\partial\mathbf{A}_{i}\ket{\Psi^{(0)}} \right\}$ contains for example $\ket{\Psi^{(0)}}$, and the second order space $\left\{ \partial^{2}/\partial\mathbf{A}_{i}\partial\mathbf{A}_{j}\ket{\Psi^{(0)}} \right\}$ contains both $\left\{ \ket{\partial_{i}\Psi^{(0)}} \right\}$ and $\ket{\Psi^{(0)}}$. In LRT, we look for independent changes orthogonal to the reference wavefunction, and therefore want to express derivative subspaces which are orthogonal to lower order derivative subspaces. In this study, we focus on a so-called non-redundant parameterization in terms of the projectors $\hat{Q}^{(n)}$, while previous studies focused on explicit expressions for the tangent space vectors in MPS terminology.\cite{haegeman2011,wouters2013,haegeman2013}

The projectors for the first order space were already introduced during the discussion of DMRG-LRT in Eqs.~\eqref{eq:realresp}--\eqref{eq:left1st}, i.e. the representation of $\hat{Q}_{i}^{(1)}$ is chosen from $\{\mathbf{Q}_{i}^{C}, \mathbf{Q}_{i}^{L}, \mathbf{Q}_{i}^{R}\}$, depending on the choice of the canonical form at site $i$. The mixed-canonical form of Eq. \eqref{eq:mps_wave} then results in an overlap matrix $\mathbf{S}_{ij}$ which is block diagonal in the site-indices. A diagonal block $\mathbf{S}_{ii}$ is of the form
\begin{equation}
  \mathbf{S}_{ii} = {\mathbf{Q}_{i}^{(1)}}^{\dag}\mathbf{Q}_{i}^{(1)}
\end{equation}
where it should be noted that the rank of the $dM^{2} \times dM^{2}$ matrix $\mathbf{S}_{ii}$ is now only $(d-1)M^{2}$, equal to the number of non-redundant parameters at site $i$, because $\mathbf{Q}_{i}^{(1)}$ explicitly projects out the zeroth order contributions.

The MPS tangent space vectors are normalized basis vectors in the span of $\mathbf{Q}_{i}^{(1)}$. The connection between the projector and non-redundant tangent space parameterization of the MPS is discussed in detail in Ref.~\onlinecite{wouters2013}. Briefly, an explicit parameterization of the non-redundant tangent space vectors is given by
\begin{equation}
  \mathcal{T}_\mathbf{A}\ket{\Psi^{(0)}}
 =\frac{1}{k} \sum_{i} \mathbf{B}_{i}\frac{\partial}{\partial\mathbf{A}_{i}}\ket{\Psi^{(0)}}
\end{equation}
where $\mathbf{B}_{i}$ describes the null-space projection $\{l_{i-1}\}\times\{n_{i}\}\rightarrow\{l_{i}\}^{\perp}$ which is given by the $dM \times (d-1)M$ matrix satisfying $\mathbf{B}_{i}^{\dag}\mathbf{B}_{i}=\mathbf{1}$
and $\mathbf{B}_{i}^{\dag}\mathbf{A}_{i}=\mathbf{0}$ for the left-fixed gauge condition.
The first order change is then parameterized by the $(d-1)M \times M$ matrix $\bm{x}_{i}$, which is related to the projector formalism by
\begin{equation}
  \mathbf{L}_{i}^{(1)} = \mathbf{B}_{i} \bm{x}_{i} \bm{\Lambda}_{i}^{-1}.
\end{equation}

For the second order derivative, the projector $\hat{Q}_{ij}^{(2)}$ cannot be simply defined for any gauge choice. However, in the mixed-canonical form, we can define the second order projector as the product of first order projectors,
$\hat{Q}_{ij}^{(2)} = \hat{Q}_{i}^{(1)} \times \hat{Q}_{j}^{(1)}$.
To check this, we investigate whether the overlap $\braket{\partial_{i}\partial_{j}\Psi^{(0)}|\partial_{p}\Psi^{(0)}}$ is zero or not in this gauge.

%%
%% DIAGRAMS TO NON-REDUNDANT PARAMETERIZATION
%%
\begin{figure}[tb]
\includegraphics[scale=0.15, bb=0 0 1368 825]{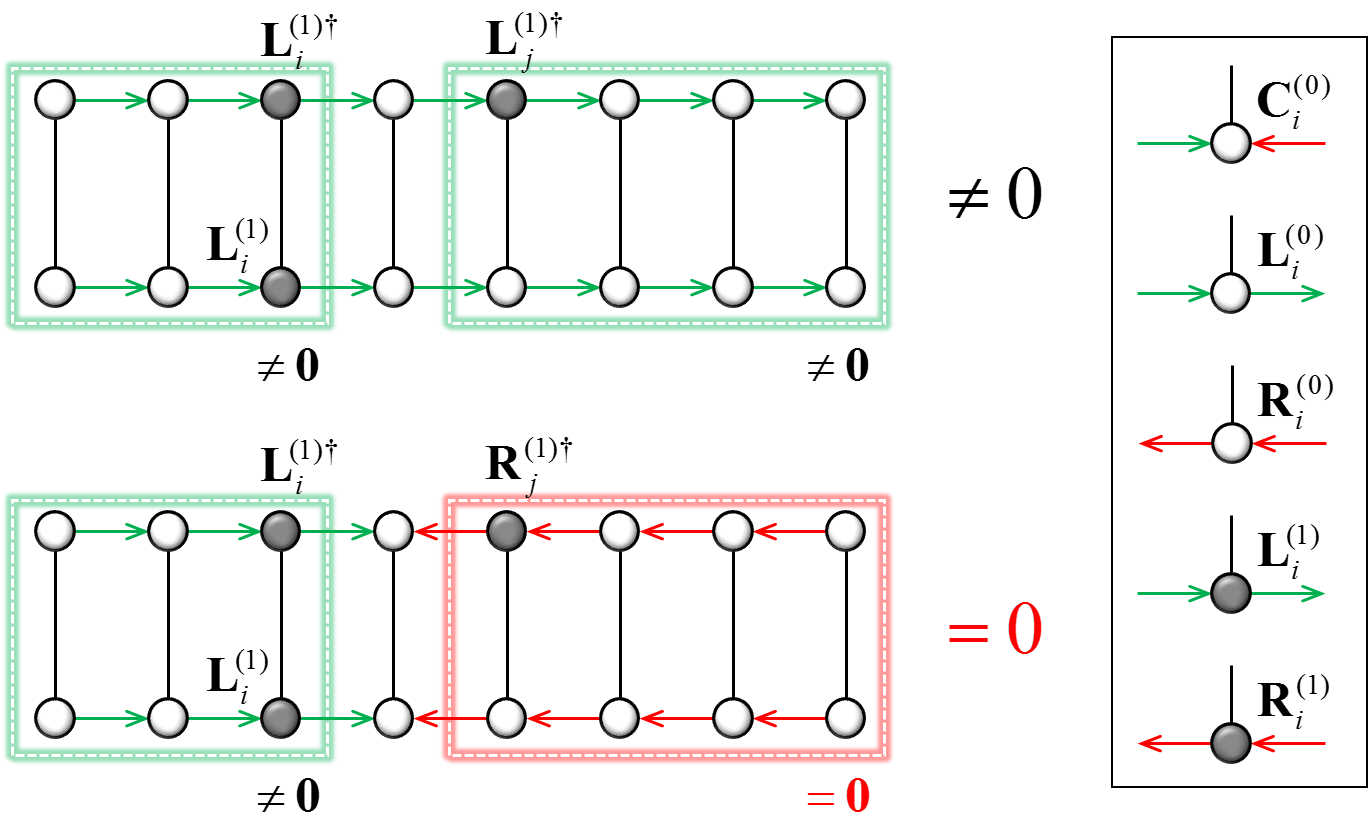}
\caption{Non-redundant parameterization for the second order derivative. The top panel shows the redundancy between the first and second derivatives with the left-canonical gauge. The bottom panel shows the non-redundant parameterization with the mixed-canonical gauge in between sites $i$ and $j$.}
\label{fig:nonred_param}
\end{figure}

If the site index of $p$ is different from the site indices of $i$ and $j$, the overlap is zero (in fact, for any gauge choice) since at least one of the first order projectors gives null. The non-trivial case occurs when the site corresponding to $p$ also corresponds to either $i$ or $j$.
%Take $p$ and $i$ to be on the same site for example, then the overlap is not zero within the left-fixed gauge condition, because the projectors at site $i$ do not ensure orthogonality between the wavefunctions anymore:
%\begin{equation}
%  \sum_{n_{j}} {\mathbf{L}^{n_{j}(1)}}^{\dag}
%  \left(\left(\sum_{n_{i}}{\mathbf{L}^{n_{i}(1)}}^{\dag}\mathbf{L}^{n_{i}(1)}\right)
%  \times \mathbb{E}_{i+1}^{j-1} \right) \mathbf{L}^{n_{j}(0)}
%  \neq \mathbf{0}
%\end{equation}
%where $\mathbb{E}_{i+1}^{j-1}$ is the transfer operator from site $i+1$ to $j-1$:
%\begin{equation}
%  \mathbb{E}_{i+1}^{j-1}
% =\prod_{p=i+1}^{j-1} \sum_{n_{p}} {\mathbf{L}^{n_{p}(0)}}^{\dag} \otimes \mathbf{L}^{n_{p}(0)}
%\end{equation}
Consider $p$ and $i$ to be on the same site within the mixed-canonical gauge, if $\mathbf{C}_{q}^{(0)}$ is in between sites $i$ and $j$, the overlap is zero because of the orthogonality in the right block, $\sum_{n_{j}} \mathbf{R}^{n_{j}(0)}{\mathbf{R}^{n_{j}(1)}}^{\dag} = \mathbf{0}$ (see Fig.~\ref{fig:nonred_param}).
%\begin{equation}
%  \left(\sum_{n_{i}}{\mathbf{L}^{n_{i}(1)}}^{\dag}\mathbf{L}^{n_{i}(1)}\right) \times
%  \mathbb{E}_{i+1}^{j-1} \times
%  \left(\sum_{n_{j}}\mathbf{R}^{n_{j}(0)}{\mathbf{R}^{n_{j}(1)}}^{\dag}\right)
% =0.
%\end{equation}
Consequently, the non-redundant parameterization of the second order space is achieved in a simple manner within the mixed-canonical form employed with the DMRG sweep.
%cannot be done simply within the fixed-gauge condition, but can be done within the mixed-canonical form.
(Note however, that for the third and higher order spaces, it seems to be impossible to simply define the projector $\hat{Q}^{(n)}$ as a product of the lower order projectors, e.g $\hat{Q}_{ijk}^{(3)}=\hat{Q}_{i}^{(1)} \times\hat{Q}_{j}^{(1)} \times \hat{Q}_{k}^{(1)}$).

%%
%% Algorithm Part I
%%

\section{Davidson Algorithm for DMRG-TDA and DMRG-RPA}
%In this section, we present an efficient algorithm within the DMRG context to solve the eigenvalue problem which appears in the DMRG-TDA and DMRG-RPA equations. Because we are usually interested in only a few excited states, the Davidson algorithm is ideal for this purpose.\cite{davidson1975}
We now present an efficient algorithm that will allow us to solve the DMRG-TDA and DMRG-RPA eigenvalue problems within the context of a DMRG sweep. The eigenvectors in the DMRG-TDA and DMRG-RPA are typically large, with $\mathcal{O}(dM^2k)$ elements. As we are usually interested in only a few excited states, the iterative Davidson algorithm is ideal for this purpose\cite{davidson1975} (Note that unlike in the ground-state DMRG, the Davidson algorithm is working here in the space of first order wavefunctions, i.e. the solutions are linear combinations of MPS). The basic idea is to formulate all the operations in the Davidson algorithm, such as the matrix-vector multiplication, in terms of a DMRG sweep.
Initially, we focus on DMRG-TDA to explain our basic algorithm, and later we explain its generalization for DMRG-RPA.

\subsection{Overall Structure of the DMRG-TDA Computations}
In the Davidson algorithm, a full matrix representation is projected into a small matrix spanned by a small number of trial vectors. In the case that several eigenvalues are desired, the block-variant of the Davidson algorithm is commonly used. Its pseudocode is shown in Fig.~\ref{alg:david_gep}.

\begin{figure}
\begin{algorithmic}[1]
  \State choose $n$ trial vectors $\mathbf{V}_{1}=\{\bm{x}_{1},\ldots,\bm{x}_{n}\}$
  \For{$i = 1,2,\ldots$}
    \State $m \gets \text{dim}(\mathbf{V}_{i})$
    \State $\mathbf{W}_{i}^{H} \gets \mathbf{H}\mathbf{V}_{i}$, $\mathbf{H}_{R} \gets \mathbf{V}_{i}^{\dag}\mathbf{W}_{i}^{H}$
    \State $\mathbf{W}_{i}^{S} \gets \mathbf{S}\mathbf{V}_{i}$, $\mathbf{S}_{R} \gets \mathbf{V}_{i}^{\dag}\mathbf{W}_{i}^{S}$
    \State solve $m$ eigenpairs $\{\omega_{\mu}, \bm{\alpha}_{\mu}\}$ of $\mathbf{H}_{R}\bm{\alpha}_{\mu} = \omega_{\mu} \mathbf{S}_{R}\bm{\alpha}_{\mu}$
    \For{$\mu = 1 \to m$}
      \State $\mathbf{u}_{\mu} \gets \mathbf{V}_{i} \bm{\alpha}_{\mu}$
      \State $\mathbf{p}_{\mu} \gets \mathbf{W}_{i}^{H} \bm{\alpha}_{\mu}$, $\mathbf{q}_{\mu} \gets \mathbf{W}_{i}^{S} \bm{\alpha}_{\mu}$
      \State $\mathbf{r}_{\mu} \gets \mathbf{p}_{\mu} - \omega_{\mu}\mathbf{q}_{\mu}$
    \EndFor
    \If{$\{|\bm{r}_{\mu}|^{2} < \epsilon : 1 \leq \mu \leq n\}$}
      \State Exit on Convergence
    \EndIf
    \For{$\mu = 1 \to n$}
      \State $\mathbf{z}_{\mu} \gets -(\mathbf{H} - \omega_{\mu}\mathbf{S})^{-1} \mathbf{r}_{\mu}$
    \EndFor
    \If{$m \leq N - n$}
      \State $\mathbf{V}_{i+1} \gets \{\mathbf{u}_{1},\ldots,\mathbf{u}_{m},\mathbf{z}_{1},\ldots,\mathbf{z}_{n}\}$
    \Else
      \State $\mathbf{V}_{i+1} \gets \{\mathbf{u}_{1},\ldots,\mathbf{u}_{n},\mathbf{z}_{1},\ldots,\mathbf{z}_{n}\}$ \Comment{Deflation}
    \EndIf
  \EndFor
\end{algorithmic}
\caption{Pseudocode of the block-Davidson algorithm for generalized eigenvalue problems (GEP),
         where $n$ is the number of desired roots, $\epsilon$ is a certain (small) threshold, and $N$ is the maximum number of trial vectors.
         Upon convergence, $\{\omega_{\mu}, \mathbf{u}_{\mu} : 1 \leq \mu \leq n\}$ are the eigenvalues and eigenvectors, respectively.}
\label{alg:david_gep}
\end{figure}

For DMRG-TDA, one particular choice $\mathbf{X}=\{\mathbf{L}_{1}^{(1)},\cdots,\mathbf{C}_{i}^{(1)},\cdots,\mathbf{R}_{k}^{(1)}\}$ of first order changes in the DMRG wavefunction is a trial vector, which has a dimension of $dM^{2}k$.
This trial vector represents formally a linear combination of MPS, of the form in Eq.~\eqref{eq:psi1st}.

Important for an efficient implementation of DMRG-TDA (and DMRG-RPA) is how to efficiently compute (1) the projected matrices $\mathbf{H}_{R}$ and $\mathbf{S}_{R}$, (2) the eigenvectors as a rotation of the basis vectors by the Ritz vectors $\bm{\alpha}$, and (3) the correction vectors $\mathbf{z}$, because they involve matrix/vector multiplications which have the large linear dimension $dM^{2}k$. A naive implementation of the matrix-vector multiplications would cost $\mathcal{O}(M^4)$. Fortunately, they can be carried out with at most $\mathcal{O}(M^{3})$ complexity by means of the sweep algorithm, as discussed in the following sections.

\subsection{Sweep algorithm for the projected Hamiltonian}
\begin{figure}[tb]
\includegraphics[scale=0.15, bb=0 0 1481 1664]{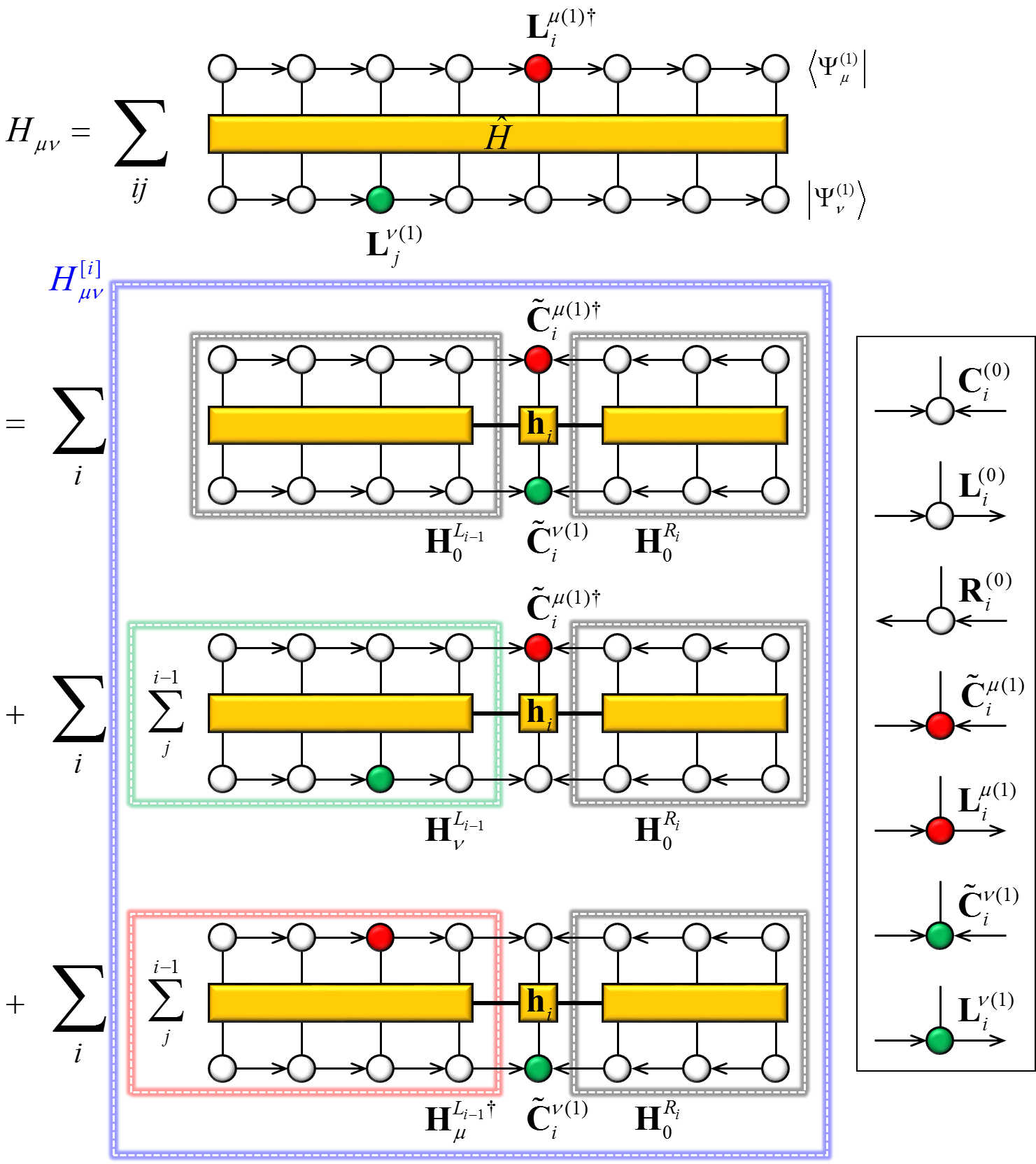}
\caption{Graphical summary for the computation of the projected Hamiltonian $H_{\mu\nu}$ in terms of the sweep algorithm.}
\label{fig:sweep_algo}
\end{figure}

Suppose we have $m$ trial vectors for the first order wavefunctions in the left-canonical form,
i.e. $\mathbf{X}_{\mu}=\{\mathbf{L}_{1}^{\mu(1)},\cdots,\mathbf{L}_{k}^{\mu(1)}\}_{1 \leq \mu \leq m}$. The matrix elements of the projected Hamiltonian $\mathbf{H}_{R}$ are then given by
\begin{equation}
\label{eq:hsub}
  H_{\mu\nu}
 =\braket{\Psi_{\mu}^{(1)}|\hat{H}|\Psi_{\nu}^{(1)}}
 =\sum_{ij} {\mathbf{L}_{i}^{\mu(1)}}^{\dag} \mathbf{H}_{ij} \mathbf{L}_{j}^{\nu(1)} .
\end{equation}
%where $\mathbf{H}_{ij}^{(0)}$ is the effective Hamiltonian spanned by the first derivative spaces for sites $i$ and $j$
%\begin{equation}
%\label{eq:heff}
%  \mathbf{H}_{ij}
% =\braket{\partial_{i}\Psi^{(0)}|\hat{H} - E_{0}|\partial_{j}\Psi^{(0)}}
%\end{equation}
%where $\ket{\partial_{i}\Psi^{(0)}}$ is not equal to the MPS tangent space because it's, in principal, redundant to zero-th order
%wavefunction and the projection onto tangent space is involved in the first order rotation matrix $\mathbf{L}_{i}^{\mu(1)}$.
To construct a sweep algorithm, it is necessary to break up the overall matrix/vector computation into site-by-site computations, such that
\begin{equation}
\label{eq:hdiv}
  H_{\mu\nu} = \sum_{i=1}^{k} H_{\mu\nu}^{[i]}
\end{equation}
%%\begin{equation}
%%\begin{split}
%%  H_{\mu\nu}
%%&=\sum_{i}   \mathbf{L}_{i}^{\mu(1)\dag} \mathbf{H}_{ii}^{(0)} \mathbf{L}_{i}^{\nu(1)} \\
%%&+\sum_{i>j} \left(
%%             \mathbf{L}_{i}^{\mu(1)\dag} \mathbf{H}_{ij}^{(0)} \mathbf{L}_{j}^{\nu(1)}
%%            +\mathbf{L}_{j}^{\mu(1)\dag} \mathbf{H}_{ji}^{(0)*} \mathbf{L}_{i}^{\nu(1)} \right)
%%\end{split}
%%\end{equation}
where $H_{\mu\nu}^{[i]}$ is defined as a (left-) block component at site~$i$
\begin{equation}
\label{eq:hleft1}
\begin{split}
  H_{\mu\nu}^{[i]}
&={\mathbf{L}_{i}^{\mu(1)}}^{\dag}\mathbf{H}_{ii}^{(0)}\mathbf{L}_{i}^{\nu(1)} \\
&+\sum_{j=1}^{i-1} \left(
  {\mathbf{L}_{i}^{\mu(1)}}^{\dag} \mathbf{H}_{ij}^{(0)} \mathbf{L}_{j}^{\nu(1)}
 +{\mathbf{L}_{j}^{\mu(1)}}^{\dag} \mathbf{H}_{ji}^{(0)} \mathbf{L}_{i}^{\nu(1)} \right)
\end{split}
\end{equation}
Taking the summation over $j$ and using the first order gauge condition in Eq.~\eqref{eq:gauge1}, we get
\begin{equation}
\label{eq:hleft2}
\begin{split}
  H_{\mu\nu}^{[i]}
&=\mbox{$\mathbf{\tilde{C}}_{i}^{\mu(1)}$}^{\dag} \mathbf{H}_{0}^{[i]} \mathbf{\tilde{C}}_{i}^{\nu(1)} \\
&+\mbox{$\mathbf{\tilde{C}}_{i}^{\mu(1)}$}^{\dag} \Delta\mathbf{H}_{\nu}^{[i]} \mathbf{C}_{i}^{(0)}
 +{\mathbf{C}_{i}^{(0)}}^{\dag} \Delta{\mathbf{H}_{\mu}^{[i]}}^{\dag} \mathbf{\tilde{C}}_{i}^{\nu(1)}
\end{split}
\end{equation}
where $\mathbf{H}_{0}^{[i]}$ and $\Delta\mathbf{H}_{\mu}^{[i]}$ are the zeroth and first order 
\textit{superblock} operators, respectively, in the left-block component
\begin{equation}
\label{eq:hzero_sb}
  \mathbf{H}_{0}^{[i]}
 =\mathbf{H}_{0}^{L_{i-1}} \times \mathbf{h}_{i} \times \mathbf{H}_{0}^{R_{i}}
\end{equation}
\begin{equation}
\label{eq:hleft_sb}
  \Delta\mathbf{H}_{\mu}^{[i]}
 =\mathbf{H}_{\mu}^{L_{i-1}} \times \mathbf{h}_{i} \times \mathbf{H}_{0}^{R_{i}}.
\end{equation}
In this equation, $\mathbf{h}_{i}$ contains the local operators acting on site $i$. $\mathbf{H}_{0}^{L_{i-1}}$ and $\mathbf{H}_{0}^{R_{i}}$ are the zeroth order left- and right-renormalized operators, respectively. $\mathbf{H}_{\mu}^{L_{i-1}}$ is the first order left-renormalized operator given by
\begin{equation}
\label{eq:lopr1st}
\begin{split}
  \mathbf{H}_{\mu}^{L_{i-1}}
&={\mathbf{L}_{i-1}^{(0)}}^{\dag} \left[ \mathbf{H}_{\mu}^{L_{i-2}} \times \mathbf{h}_{i-1} \right] \mathbf{L}_{i-1}^{(0)} \\
&+{\mathbf{L}_{i-1}^{(0)}}^{\dag} \left[ \mathbf{H}_{0}^{L_{i-2}} \times \mathbf{h}_{i-1}   \right] \mathbf{L}_{i-1}^{\mu(1)}
\end{split}
\end{equation}
where the summation over complementary operators in the DMRG formalism (or the contraction of virtual bonds in the MPO formalism) is abbreviated as a multiplication symbol ($\times$).
These equations are summarized diagrammatically in Fig.~\ref{fig:sweep_algo}.

%%
%% SWEEP DIAGRAMS
%%

In Eq.~\eqref{eq:hleft2}, it should be noted that we use the projected coefficient
\begin{equation}
\label{eq:cwfn1st}
  \mathbf{\tilde{C}}_{i}^{\mu(1)}
 =\mathbf{Q}_{i}^{L}\mathbf{C}_{i}^{\mu(1)}
 =\left(\mathbf{1} - \mathbf{L}_{i}^{(0)}{\mathbf{L}_{i}^{(0)}}^{\dag}\right)\mathbf{C}_{i}^{\mu(1)}
\end{equation}
to satisfy the left-fixed gauge condition of the first order wavefunction.

Similarly, the matrix elements of the projected overlap matrix $\mathbf{S}_{R}$ can be computed as
\begin{equation}
\label{eq:sleft2}
  S_{\mu\nu}
 =\sum_{i=1}^{k} S_{\mu\nu}^{[i]}
 =\sum_{i=1}^{k} \mbox{$\mathbf{\tilde{C}}_{i}^{\mu(1)}$}^{\dag}\mathbf{\tilde{C}}_{i}^{\nu(1)}
\end{equation}
Note that any inter-site components of the overlap matrix vanish since ${\mathbf{L}_{i}^{\mu(1)}}^{\dag}\mathbf{L}_{i}^{(0)}=\mathbf{0}$.

\subsection{Rotation of the First Order Vectors by the Ritz Vectors}
At the end of the sweep, we have the small generalized eigenvalue problem to solve:
\begin{equation}
\label{eq:eigs}
  \mathbf{H}_{R} \bm{\alpha}_{\mu} = \omega_{\mu} \mathbf{S}_{R} \bm{\alpha}_{\mu}.
\end{equation}
This gives an approximation of the excitation energy $\omega_{\mu}$ and the Ritz vector $\bm{\alpha}_{\mu}$.
The Ritz vector is then used to rotate the first order objects to approximate the first order eigenstates:
\begin{equation}
\label{eq:ritzrot1}
\begin{split}
  \mathbf{L}_{i}^{\mu(1)}
&=\sum_{\mu'} \mathbf{L}_{i}^{\mu'(1)} \alpha_{\mu'\mu} \\
  \mathbf{H}_{\mu}^{L_{i-1}}
&=\sum_{\mu'} \mathbf{H}_{\mu'}^{L_{i-1}} \alpha_{\mu'\mu}
\end{split}
\end{equation}
where $\mu'$ denotes the state index in the previous sweep. Note that the rotation performs the orthonormalization of trial vectors as well.
Because the explicit orthonormalization of trial vectors, in principle, takes another sweep, it is advantageous to solve
the small generalized eigenvalue problem so that the rotation and the orthonormalization can be done simultaneously.

\begin{figure}[tb]
\includegraphics[scale=0.15, bb=0 0 1228 973]{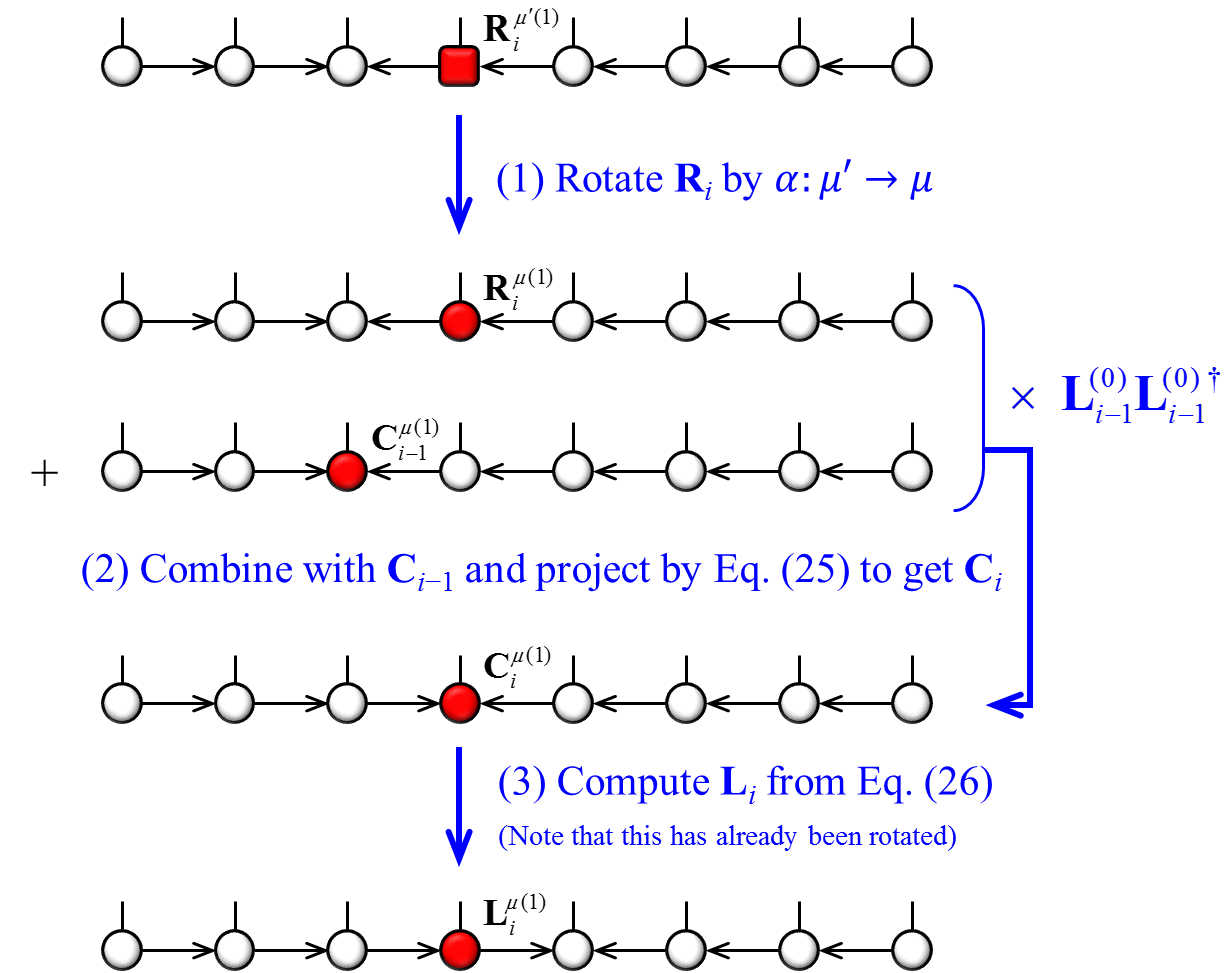}
\caption{Graphical summary for the Ritz vector rotation, gauge-transformation, and computation of the first order rotation matrix, through sweeping.}
\label{fig:gauge_trans}
\end{figure}

As in the ground-state DMRG, sweeps are arranged as successive forward and backward iterations.
In the forward sweep, for example, we only need to rotate the right-block objects (and vice versa in the backward sweep), i.e.
\begin{equation}
\label{eq:ritzrot2}
\begin{split}
  \mathbf{R}_{i}^{\mu(1)}
&=\sum_{\mu'} \mathbf{R}_{i}^{\mu'(1)} \alpha_{\mu'\mu} \\
  \mathbf{H}_{\mu}^{R_{i}}
&=\sum_{\mu'} \mathbf{H}_{\mu'}^{R_{i}} \alpha_{\mu'\mu}
%&\mathbf{C}_{i}^{\mu(1)}
%=\mathbf{L}_{i-1}^{(0)\dag}\mathbf{C}_{i-1}^{\mu(1)}\mathbf{R}_{i}^{(0)}
%+\mathbf{L}_{i-1}^{(0)\dag}\mathbf{C}_{i-1}^{(0)}\mathbf{R}_{i}^{\mu(1)}
\end{split}
\end{equation}
because the left-block objects such as $\mathbf{L}_{i-1}^{\mu(1)}$ and $\mathbf{H}_{\mu}^{L_{i-1}}$ have already been rotated
at the previous site.
%during the previous backward sweep.
These steps are illustrated diagrammatically, in Fig.~\ref{fig:gauge_trans}.

%%
%% DAVIDSON SWEEP
%%

%The \textit{left-block} formulae are used to carry out a forward sweep and
%the \textit{right-block} formulae can also be derived in the same way to carry out a backward sweep.
%Thus, one step of Davidson's algorithm for DMRG-TDA is performed with one sweep iteration. 

\subsection{Davidson's correction equation}
To compute the correction site-by-site, we consider the $i$-th site component of the sigma-vector in the mixed-canonical gauge
\begin{equation}
\label{eq:sigv1}
  \bm{\sigma}_{i}^{\mu}
 =\mathbf{H}_{0}^{[i]} \mathbf{C}_{i}^{\mu(1)} + \Delta\bm{\sigma}_{\mu}^{[i]} \mathbf{C}_{i}^{(0)}
\end{equation}
\begin{equation}
\label{eq:sigv2}
\begin{split}
  \Delta\bm{\sigma}_{\mu}^{[i]}
&=\mathbf{H}_{\mu}^{L_{i-1}} \times \mathbf{h}_{i} \times \mathbf{H}_{0}^{R_{i}} \\
&+\mathbf{H}_{0}^{L_{i-1}} \times \mathbf{h}_{i} \times \mathbf{H}_{\mu}^{R_{i}}
\end{split}
\end{equation}
where the first order right-renormalized operator $\mathbf{H}_{\mu}^{R_{i}}$ is given similarly to Eq.~\eqref{eq:lopr1st}
\begin{equation}
\label{eq:ropr1st}
\begin{split}
  \mathbf{H}_{\mu}^{R_{i}}
&=\mathbf{R}_{i+1}^{(0)} \left[ \mathbf{H}_{\mu}^{R_{i+1}} \times \mathbf{h}_{i+1} \right] {\mathbf{R}_{i+1}^{(0)}}^{\dag} \\
&+\mathbf{R}_{i+1}^{(0)} \left[ \mathbf{H}_{0}^{R_{i+1}} \times \mathbf{h}_{i+1}   \right] {\mathbf{R}_{i+1}^{\mu(1)}}^{\dag}.
\end{split}
\end{equation}

The residual vector for each site $\mathbf{r}_{i}^{\mu}$ can then be computed as
\begin{equation}
\label{eq:rnorm}
  \mathbf{r}_{i}^{\mu}=\bm{\sigma}_{i}^{\mu}-\omega_{\mu} \mathbf{C}_{i}^{\mu(1)}
\end{equation}
where we have assumed that the trial vectors are already rotated by the Ritz vectors. Note that $\mathbf{S}_{ij}^{(0)}=\mathbf{0}$ for $i \neq j$ and $\mathbf{S}_{ii}^{(0)}=\mathbf{1}$ in the mixed-canonical gauge at site~$i$.

The correction for each site $\mathbf{z}_{i}$ is then computed as
\begin{equation}
\label{eq:error}
  \mathbf{z}_{i}^{\mu}
 =\mathbf{r}_{i}^{\mu} \left[ \text{diag.}(\mathbf{H}_{0}^{[i]} - \omega_{\mu}\mathbf{1}) \right]^{-1}
\end{equation}
in which the diagonal preconditioner is employed. Note that the diagonal block of the effective Hamiltonian $\mathbf{H}_{ii}^{(0)}$, in the mixed-canonical gauge, is equal to the zeroth order Hamiltonian $\mathbf{H}_{0}^{[i]}$ for the DMRG optimization.

Because $\mathbf{z}_{i}^{\mu}$ does not lie entirely in the nullspace of the zeroth order component, it is projected into the first order orthogonal subspace to yield a new trial vector:
\begin{equation}
\label{eq:param}
  \mathbf{C}_{i}^{m+\mu(1)}
 =\left( \mathbf{1} - \mathbf{C}_{i}^{(0)} {\mathbf{C}_{i}^{(0)}}^{\dag} \right) \mathbf{z}_{i}^{\mu}.
\end{equation}
After this step, we continue the algorithm by approximating eigenstates in the larger trial space.

\subsection{Generalization to DMRG-RPA}
DMRG-RPA can be implemented with relatively minor modifications of DMRG-TDA.

Suppose we have $m$ trial vectors $\ket{\mathbf{X}^{\mu},\mathbf{Y}^{\mu}}$ with the approximate eigenvalues $\omega_{\mu}$. If these frequencies differ from zero, the vectors $\ket{\mathbf{Y}^{\mu},\mathbf{X}^{\mu}}$ are also solutions, with the eigenvalues $-\omega_{\mu}$, due to the completeness of the RPA equation. The DMRG-RPA problem is then reduced to a $2m \times 2m$ non-hermitian eigenvalue problem spanned by the vectors $\{\ket{\mathbf{X}^{\mu},\mathbf{Y}^{\mu}},\ket{\mathbf{Y}^{\mu},\mathbf{X}^{\mu}}\}$.\cite{olsen1988}

The projected DMRG-RPA equation \eqref{eq:dmrgrpa} is given by
\begin{equation}
\label{eq:reducedrpa}
\begin{pmatrix}
\mathbf{H}_{R}     & \mathbf{W}_{R}     \\
\mathbf{W}_{R}^{*} & \mathbf{H}_{R}^{*} \\
\end{pmatrix}
\begin{pmatrix}
\bm{\alpha}_{X} \\
\bm{\alpha}_{Y} \\
\end{pmatrix}
=
\omega
\begin{pmatrix}
 \mathbf{S}_{R} &  \mathbf{D}_{R}     \\
-\mathbf{D}_{R}^{*} & -\mathbf{S}_{R}^{*} \\
\end{pmatrix}
\begin{pmatrix}
\bm{\alpha}_{X} \\
\bm{\alpha}_{Y} \\
\end{pmatrix}
\end{equation}
in terms of the following $m \times m$ matrices:
\begin{align}
  H_{\mu\nu}
&=\begin{pmatrix}
        \mathbf{X}^{\mu} & \mathbf{Y}^{\mu}
  \end{pmatrix}
  \begin{pmatrix}
        \mathbf{H}     & \mathbf{W} \\
        \mathbf{W}^{*} & \mathbf{H}^{*} \\
  \end{pmatrix}
  \begin{pmatrix}
        \mathbf{X}^{\nu} \\
        \mathbf{Y}^{\nu} \\
  \end{pmatrix} \\
  W_{\mu\nu}
&=\begin{pmatrix}
        \mathbf{X}^{\mu} & \mathbf{Y}^{\mu}
  \end{pmatrix}
  \begin{pmatrix}
        \mathbf{H}     & \mathbf{W} \\
        \mathbf{W}^{*} & \mathbf{H}^{*} \\
  \end{pmatrix}
  \begin{pmatrix}
        \mathbf{Y}^{\nu} \\
        \mathbf{X}^{\nu} \\
  \end{pmatrix} \\
  S_{\mu\nu}
&=\begin{pmatrix}
        \mathbf{X}^{\mu} & \mathbf{Y}^{\mu}
  \end{pmatrix}
  \begin{pmatrix}
        \mathbf{S} &  \mathbf{0} \\
        \mathbf{0} & -\mathbf{S}^{*} \\
  \end{pmatrix}
  \begin{pmatrix}
        \mathbf{X}^{\nu} \\
        \mathbf{Y}^{\nu} \\
  \end{pmatrix} \\
  D_{\mu\nu}
&=\begin{pmatrix}
        \mathbf{X}^{\mu} & \mathbf{Y}^{\mu}
  \end{pmatrix}
  \begin{pmatrix}
        \mathbf{S} &  \mathbf{0} \\
        \mathbf{0} & -\mathbf{S}^{*} \\
  \end{pmatrix}
  \begin{pmatrix}
        \mathbf{Y}^{\nu} \\
        \mathbf{X}^{\nu} \\
  \end{pmatrix}.
\end{align}
%To solve a generalized non-hermitian eigenvalue problem, it is convenient to rewrite Eq.~(\ref{eq:reducedrpa}) as
%\begin{equation}
%\label{eq:invmetricrpa}
%\frac{1}{\omega}
%\begin{pmatrix}
%\mathbf{H}_{R}     & \mathbf{W}_{R}     \\
%\mathbf{W}_{R}^{*} & \mathbf{H}_{R}^{*} \\
%\end{pmatrix}
%\begin{pmatrix}
%\bm{\tilde\alpha}_{X} \\
%\bm{\tilde\alpha}_{Y} \\
%\end{pmatrix}
%=
%\begin{pmatrix}
% \mathbf{S}_{R}     &  \mathbf{D}_{R}     \\
%-\mathbf{D}_{R}^{*} & -\mathbf{S}_{R}^{*} \\
%\end{pmatrix}
%\begin{pmatrix}
%\bm{\tilde\alpha}_{X} \\
%\bm{\tilde\alpha}_{Y} \\
%\end{pmatrix}
%\end{equation}
%where the stability matrix
%$\left( \begin{smallmatrix} \mathbf{H}_{R} & \mathbf{W}_{R} \\ \mathbf{W}_{R}^{*} & \mathbf{H}_{R}^{*} %\end{smallmatrix} \right)$
%is considered as a metric and the original eigenvector of the reduced RPA equation is of the form
%\begin{equation}
%  \begin{pmatrix} \bm{\alpha}_{X} \\ \bm{\alpha}_{Y} \end{pmatrix}
% =\frac{1}{\sqrt{\omega}}
%  \begin{pmatrix} \bm{\tilde\alpha}_{X} \\ \bm{\tilde\alpha}_{Y} \end{pmatrix}
%\end{equation}
The rotation of the first order vectors by the Ritz vectors has to be carried out for the forward and backward propagating parts $\{\bm{\alpha}_{X},\bm{\alpha}_{Y}\}$ simultaneously:
\begin{equation}
\begin{split}
  \mathbf{X}_{i}^{\mu}
&=\sum_{\mu'} \left( \mathbf{X}_{i}^{\mu'} \alpha_{X,\mu'\mu} + \mathbf{Y}_{i}^{\mu'} \alpha_{Y,\mu'\mu} \right) \\
  \mathbf{Y}_{i}^{\mu}
&=\sum_{\mu'} \left( \mathbf{Y}_{i}^{\mu'} \alpha_{X,\mu'\mu} + \mathbf{X}_{i}^{\mu'} \alpha_{Y,\mu'\mu} \right).
\end{split}
\end{equation}

To compute the second order derivative contributions in the $\mathbf{W}$ matrix, the non-redundant parameterization is performed by taking the projection $\mathbf{Q}_{i}^{L} \times \mathbf{Q}_{j}^{R} : (i < j)$, as discussed before.

For example, the non-redundant contribution coming from a pair of nearest neighbour sites $\mathbf{C}_{i}^{(1)}\mathbf{R}_{i+1}^{(1)}$ is obtained as
\begin{equation}
  \mathbf{Q}_{i}^{L}\mathbf{C}_{i}^{(1)}\mathbf{R}_{i+1}^{(1)}\mathbf{Q}_{i+1}^{R}
 =\mathbf{L}_{i}^{(1)}\bm{\Lambda}_{i}\mathbf{R}_{i+1}^{(1)}.
 \end{equation}
For the next nearest neighbour sites as
\begin{equation}
  \mathbf{Q}_{i-1}^{L}\mathbf{C}_{i}^{(1)}\mathbf{R}_{i}^{(0)}\mathbf{R}_{i+1}^{(1)}\mathbf{Q}_{i+1}^{R}
 =\mathbf{L}_{i-1}^{(1)}\mathbf{C}_{i}^{(0)}\mathbf{R}_{i+1}^{(1)}
\end{equation}
and so forth.

\begin{figure}[tb]
\includegraphics[scale=0.15, bb=0 0 1268 917]{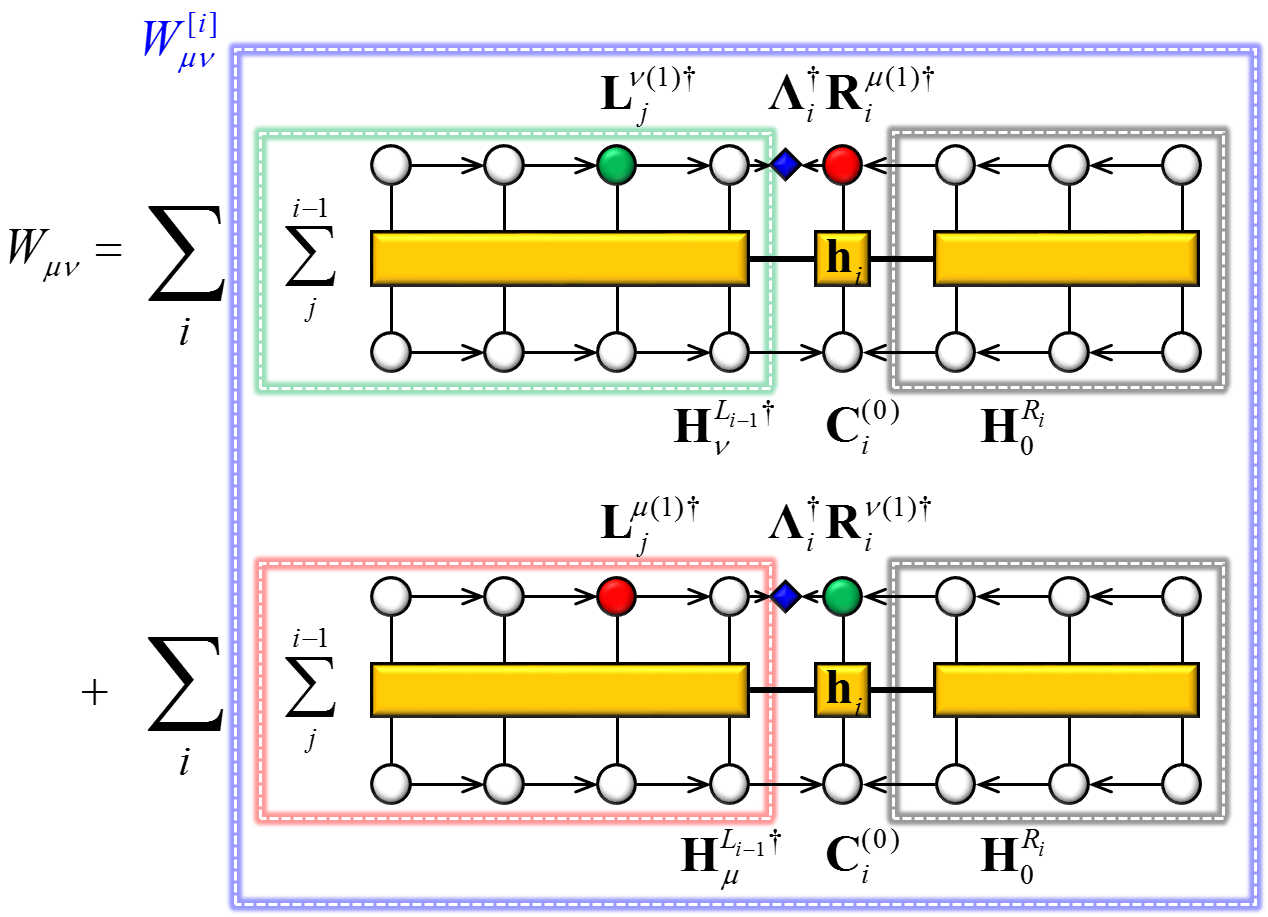}
\caption{Graphical representation for computation of $W_{\mu\nu}^{[i]}$.}
\label{fig:compt_wmat}
\end{figure}

%To take these second order changes into account, the matrix elements $W_{\mu\nu}^{[i]}$ are computed as
%\begin{equation}
%\label{eq:bmatrix}
%  W_{\mu\nu}^{[i]}
% =(\bm{\Lambda}_{i}\mathbf{R}_{i}^{\mu(1)})^{\dag}\Delta{\mathbf{H}_{\nu}^{[i]}}^{\dag} \mathbf{C}_{i}^{(0)}
%\end{equation}
To take these second order changes into account, the matrix elements $W_{\mu\nu}$ are computed as
\begin{equation}
\label{eq:bmatrix}
\begin{split}
  W_{\mu\nu}
&=\sum_{i} W_{\mu\nu}^{[i]} \\
&=\sum_{i} (\bm{\Lambda}_{i}\mathbf{R}_{i}^{\mu(1)})^{\dag}\Delta{\mathbf{H}_{\nu}^{[i]}}^{\dag} \mathbf{C}_{i}^{(0)}
          +(\bm{\Lambda}_{i}\mathbf{R}_{i}^{\nu(1)})^{\dag}\Delta{\mathbf{H}_{\mu}^{[i]}}^{\dag} \mathbf{C}_{i}^{(0)}
\end{split}
\end{equation}
where $\Delta\mathbf{H}_{\nu}^{[i]}$ is the same as Eq.~\eqref{eq:hleft_sb}, but now the conjugate is considered. Note that the matrix elements of $\mathbf{W}$ have to be computed through two sweeps
because they depend on both $\mathbf{L}_{i}^{(1)}$ and $\mathbf{R}_{i}^{(1)}$, which are obtained from forward and backward sweeps, respectively.
The computation of $W_{\mu\nu}$ is illustrated diagrammatically in Fig.~\ref{fig:compt_wmat}

%%
%% W COMPUTATION
%%

%%
%% Pseudo Code
%%

\section{Pseudocode for DMRG-TDA/RPA}
In Fig.~\ref{alg:dmrgtda}, we summarize our DMRG-TDA algorithm to target the $n$ lowest excited states as a pseudocode.
All zeroth order information, $\{\mathbf{L}_{i}^{(0)}$, $\mathbf{C}_{i}^{(0)}$, $\mathbf{R}_{i}^{(0)}\}$, $\mathbf{H}_{i}^{L_{i}}$ and $\mathbf{H}_{i}^{R_{i}}$, is supposed to be known from the ground-state DMRG calculation. The overall complexity of our algorithm is $\mathcal{O}(dM^{3}k^{3}N+dM^{2}kN^{2})$, where N is the maximum number of trial vectors in the Davidson algorithm. We have implemented these DMRG-TDA and DMRG-RPA algorithms into our spin-adapted DMRG code (\textsc{Block}).\cite{chan2002, sharma2011, sharma2012}

\begin{figure*}
%\begin{subfigure}[t]{0.47\hsize}
\begin{minipage}[t]{0.47\hsize}
\begin{algorithmic}[1]
\State choose $n$ trial vectors $\{\mathbf{X}_{\mu} : 1 \leq \mu \leq n\}$
\State $m \gets n$
\Loop \Comment{Iteration for the Davidson Algorithm}
  \State solve $\mathbf{H}_{R}\bm{\alpha}_{\mu}=\omega_{\mu}\mathbf{S}_{R}\bm{\alpha}_{\mu}$
  \State $l \gets m + n$
  \State $H_{\mu\nu} \gets 0, S_{\mu\nu} \gets 0 : \mu,\nu = 1,\ldots,l$
  \State $r_{\mu}^{2} \gets 0$
  \For{$i = 1 \to k$} \Comment{Start Forward Sweep}
    \For{$\mu = 1 \to m$}
%\Statex{Rotation by $\bm{\alpha}$}
      \State $\mathbf{R}_{i}^{\mu(1)},\mathbf{H}_{\mu}^{R_{i}} \gets$ Eq.~\eqref{eq:ritzrot2}
      \State $\mathbf{C}_{i}^{\mu(1)} \gets$ Eq.~\eqref{eq:gauge_fwd}
    \EndFor
%\Statex{Correction Equation}
    \For{$\mu = 1 \to n$}
      \State $\bm{\sigma}_{i}^{\mu} \gets$ Eq.~\eqref{eq:sigv1}
      \State $\mathbf{r}_{i}^{\mu} \gets$ Eq.~\eqref{eq:rnorm}
      \State $\mathbf{C}_{i}^{m+\mu(1)} \gets$ Eq.~\eqref{eq:param}
      \State $r_{\mu}^{2} \gets r_{\mu}^{2} + |\mathbf{r}_{i}^{\mu}|^{2}$
    \EndFor
%\Statex{Projected Coefficient}
    \For{$\mu = 1 \to l$}
      \State $\mathbf{\tilde{C}}_{i}^{\mu(1)} \gets$ Eq.~\eqref{eq:cwfn1st}
    \EndFor
%\Statex{Reduced Matrix Elements}
    \For{$\mu = 1 \to l$}
      \For{$\nu = 1 \to l$}
        \State $H_{\mu\nu} \gets H_{\mu\nu} + H_{\mu\nu}^{[i]}$\Comment{Eq.~\eqref{eq:hleft2}}
        \State $S_{\mu\nu} \gets S_{\mu\nu} + S_{\mu\nu}^{[i]}$\Comment{Eq.~\eqref{eq:sleft2}}
      \EndFor
    \EndFor
\algstore{bkbreak}
\end{algorithmic}
%\end{subfigure}
\end{minipage}
~~
%\begin{subfigure}[t]{0.47\hsize}
\begin{minipage}[t]{0.47\hsize}
\begin{algorithmic}[1]
\algrestore{bkbreak}
%\Statex{Renormalization}
    \For{$\mu = 1 \to l$}
      \State $\mathbf{L}_{i}^{\mu(1)} \gets$ Eq.~\eqref{eq:left1st}
      \State $\mathbf{H}_{\mu}^{L_{i}} \gets$ Eq.~\eqref{eq:lopr1st}
    \EndFor
  \EndFor \Comment{End Forward Sweep}
  \For{$i = k \to 1$} \Comment{Start Backward Sweep}
%\Statex{Renormalization}
    \For{$\mu = 1 \to l$}
      \State $\mathbf{R}_{i}^{\mu(1)} \gets$ Eq.~\eqref{eq:right1st}
      \State $\mathbf{H}_{\mu}^{R_{i-1}} \gets$ Eq.~\eqref{eq:ropr1st}
    \EndFor
%   \If{RPA} \Comment{hack here for the DMRG-RPA}
%     \State $W_{\mu\nu}^{[i]} \gets$ Eq.~(\ref{eq:bmatrix})
%   \EndIf
  \EndFor \Comment{End Backward Sweep}
  \If{$\{|\mathbf{r}_{\mu}|^{2} < \epsilon : 1 \leq \mu \leq n\}$}
    \State Exit on Convergence
  \EndIf
  \If{$l < N$}
    \State $m \gets l$
  \Else
    \State $m \gets n$ \Comment{Deflation}
  \EndIf
\EndLoop
\end{algorithmic}
%\end{subfigure}
\end{minipage}
\caption{Pseudocode for the DMRG-TDA calculation of the $n$ lowest excited states. The zeroth order components $\mathbf{L}_{i}^{(0)}$, $\mathbf{C}_{i}^{(0)}$, and $\mathbf{R}_{i}^{(0)}$ are assumed to be solved and stored. The $\mu$-th trial vector $\mathbf{X}_{\mu}$ consists of a set of first order changes $\mathbf{L}_{i}^{\mu(1)}$, $\mathbf{C}_{i}^{\mu(1)}$, and $\mathbf{R}_{i}^{\mu(1)}$. The first order right-block objects, $\mathbf{H}_{R}$ and $\mathbf{S}_{R}$, are assumed to be computed for the first $n$ trial vectors before the algorithm starts. The initial trial vectors are random. The real (and small) number $\epsilon$ gives the threshold for convergence.}
%         before carry out the DMRG-TDA calculation. These can be computed once calling the program with which $m=0$, $\bm{\alpha}=1$,
%         and $\mathbf{C}_{i}^{\mu(1)}$ is set to be a random vector or something better. Real number $\epsilon$ gives the threshold for convergence.}
\label{alg:dmrgtda}
\end{figure*}

\section{Qualitative understanding of DMRG-LRT excitations}

Before proceeding to numerical applications, we briefly discuss the physics
of  excitations that appear in DMRG-TDA and DMRG-RPA. It is not immediately straightforward to link site-based 
excitation theories (such as DMRG-TDA and DMRG-RPA)
to standard chemical intuition, which is formulated in terms of particle-based excitations. 
A site-based excitation hierarchy will efficiently capture certain kinds of high-particle rank excitations (that
are difficult to describe in particle-based theories), while poorly describing other
kinds of excitations with low-particle rank character.
While both DMRG-TDA and DMRG-RPA are formally exact as $M\to\infty$, the rate of convergence 
for different kinds of excited states may nonetheless be very different.

Generically, we can write an exact excited state $\ket{\Psi'}$ as an  operator $\hat{\Omega}$ 
acting on the exact ground-state $\ket{\Psi_0}$,
\begin{align}
\ket{\Psi'} = \hat{\Omega} \ket{\Psi_0} = \sum_i c_i \hat{O}_i \ket{\Psi_0}
\end{align}
If one approximates $\ket{\Psi_0}$ by an approximate ground state, this gives rise to an ansatz for excited states with a long history,
%known historically variously as the Feynman--Bijl ansatz,\cite{bijl1941,feynman1954} and the single mode approximation.\cite{auerbach1998}
known historically both as the Feynman--Bijl ansatz,\cite{bijl1941,feynman1954} and the single mode approximation.\cite{auerbach1998}
In quantum chemistry this is the basis of ``internally contracted'' methods for excitations,
such as internally contracted multi-reference configuration interaction, and equation of motion coupled cluster,
where $\hat{O}_i$ are the excitation operators.

An approximate particle-based excitation theory, such as EOM-CCSD, based
on a Hartree-Fock ground-state $\ket{\Phi_0}$, performs well if $\hat{O}_i$ are low-rank 
particle operators (such as single-particle excitation operators) {\it and} if $|| \hat{O}_i (\ket{\Psi_0} - \ket{\Phi_0})||$ is small. The
latter is not true if $\hat{O}_i$  annihilates $\ket{\Phi_0}$, which will happen if $\hat{O}_i$ is a virtual to
virtual excitation, since virtuals are not occupied in the ground-state. Such excited states must then be
described using a higher-rank excitation operator, even if the {\it exact} excitation operator $\hat{\Omega}$ is in fact
low-rank.

DMRG-TDA and DMRG-RPA on the other hand, perform well when the $\hat{O}_i$ are operators that act only on a small
number of neighbouring sites on the DMRG mapping to a 1D lattice, as discussed extensively
in Refs~\onlinecite{PhysRevB.85.100408}, \onlinecite{haegeman2013}, \onlinecite{haegeman2013prl}.
Since multi-particle operators (such as $\hat{n}_\alpha \hat{n}_\beta$)
can be defined even for a single site, this means that certain kinds of multi-particle excitations may be
efficiently captured.

Singly excited states can be obtained using an excitation of the form $\sum_{ij} C_{ij} \hat{a}^\dag_i \hat{a}_j$ where $i,j$ denote local sites.
This implies that DMRG-TDA and DMRG-RPA are most efficient when the bandwidth of $C_{ij}$ is small, which is the case for tightly bound excitons.
Similarly $\hat{O}_i$ should not annihilate the approximate DMRG ground-state, which can happen if the approximate ground-state does not populate
some quantum number sectors at certain lattice partitions, which are involved in the exact excitation operator
(similar to certain orbitals not being occupied in the ground state). We will see examples of this in the calculations that follow.

%%
%% Benchmark Calculations
%%

%%
%% Table: M-sweep for C08H10
%%
\begin{table*}[tb]
\caption{Energy errors in m$E_{h}$ for the C$_{8}$H$_{10}$ molecule, computed by 8SA-DMRG and DMRG-TDA with $M$ = 50, 100, 150, and 200. The converged energy $E_{\text{conv.}}$ is computed by an 8SA-DMRG calculation with $M$ = 1000.}
\label{table:m_sweep08}
\begin{ruledtabular}
\begin{tabular}{ c c c c c c c c c c }

\multirow{2}{*}{state} & \multirow{2}{*}{$E_{\text{conv.}}$ / $E_{h}$} & \multicolumn{2}{c}{$M$ =  50} & \multicolumn{2}{c}{$M$ = 100} & \multicolumn{2}{c}{$M$ = 150} & \multicolumn{2}{c}{$M$ = 200} \\
\cline{3-4}\cline{5-6}\cline{7-8}\cline{9-10}
          &               &  SA    &  TDA   &  SA    &  TDA   &  SA    &  TDA   &  SA    &  TDA   \\
\hline
XA$_{g}$  &  -308.839603  &  0.28  &  0.01  &  0.05  &  0.00  &  0.01  &  0.00  &  0.00  &  0.00  \\
2A$_{g}$  &  -308.662869  &  1.58  &  1.13  &  0.11  &  0.12  &  0.02  &  0.07  &  0.01  &  0.03  \\
1B$_{u}$  &  -308.621251  &  1.21  &  0.41  &  0.11  &  0.07  &  0.02  &  0.04  &  0.01  &  0.01  \\
2B$_{u}$  &  -308.610083  &  1.17  &  0.22  &  0.17  &  0.03  &  0.03  &  0.01  &  0.01  &  0.00  \\
3A$_{g}$  &  -308.597039  &  0.94  &  0.51  &  0.11  &  0.10  &  0.02  &  0.05  &  0.01  &  0.03  \\
4A$_{g}$  &  -308.560117  &  2.81  & 14.90  &  0.15  &  2.23  &  0.04  &  1.15  &  0.01  &  0.62  \\
3B$_{u}$  &  -308.534448  &  1.71  &  1.63  &  0.14  &  0.25  &  0.03  &  0.14  &  0.01  &  0.13  \\
5A$_{g}$  &  -308.528264  &  1.02  &  0.13  &  0.14  &  0.04  &  0.03  &  0.03  &  0.01  &  0.03  \\

\end{tabular}
\end{ruledtabular}
\end{table*}

\begin{figure*}[tb]
\includegraphics[scale=0.25, bb=0 0 1943 608]{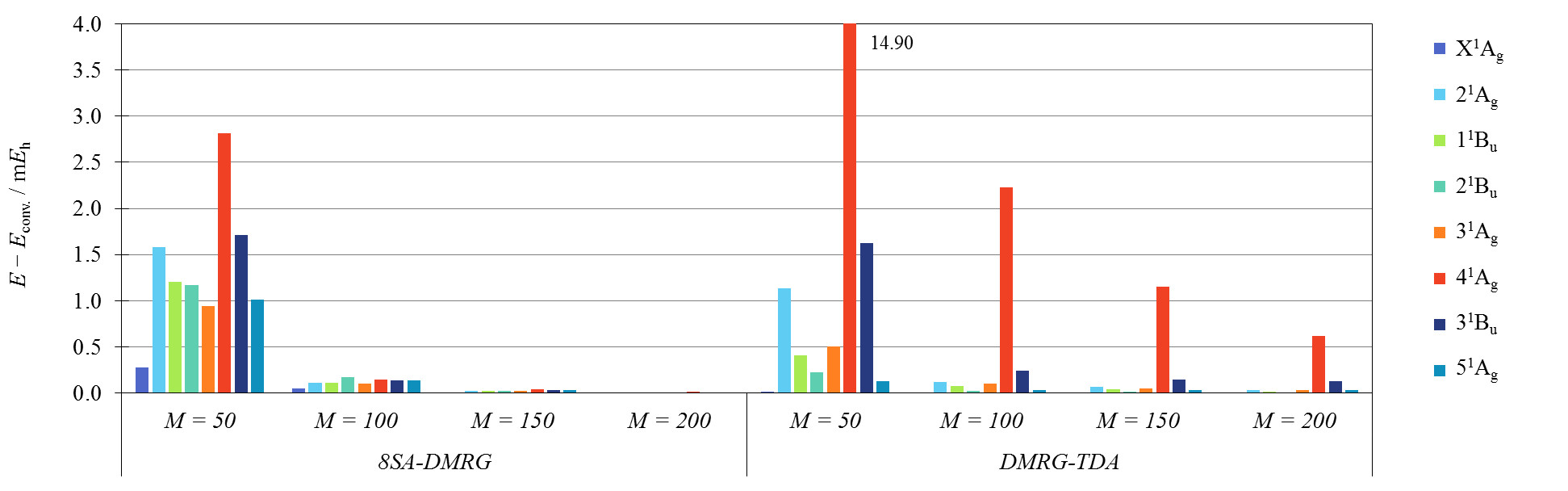}
\caption{Energy errors in m$E_{h}$ for the C$_{8}$H$_{10}$ molecule; a graphical summary for Table \ref{table:m_sweep08}.}
%\caption{Energy errors in m$E_{h}$ for the C$_{8}$H$_{10}$ molecule, computed by 8SA-DMRG and DMRG-TDA with $M$ = 50, 100, 150, and 200. The converged energy $E_{\text{conv.}}$ is computed by an 8SA-DMRG calculation with $M$ = 1000.}
\label{fig:m_sweep08}
\end{figure*}

%%
%% Table: M-sweep for C16H18
%%
\begin{table*}[tb]
\caption{Energy errors in m$E_{h}$ for the C$_{16}$H$_{18}$ molecule, computed by 8SA-DMRG and DMRG-TDA with $M$ = 50, 100, 150, and 200. The converged energy $E_{\text{conv.}}$ is computed by an 8SA-DMRG calculation with $M$ = 1000.}
\label{table:m_sweep16}
\begin{ruledtabular}
\begin{tabular}{ c c c c c c c c c c }

\multirow{2}{*}{state} & \multirow{2}{*}{$E_{\text{conv.}}$ / $E_{h}$} & \multicolumn{2}{c}{$M$ =  50} & \multicolumn{2}{c}{$M$ = 100} & \multicolumn{2}{c}{$M$ = 150} & \multicolumn{2}{c}{$M$ = 200} \\
\cline{3-4}\cline{5-6}\cline{7-8}\cline{9-10}
          &               &  SA    &  TDA   &  SA    &  TDA   &  SA    &  TDA   &  SA    &  TDA   \\
\hline
XA$_{g}$  &  -616.536393  &  2.24  &  0.13  &  0.56  &  0.02  &  0.21  &  0.01  &  0.09  &  0.00  \\
2A$_{g}$  &  -616.415267  & 10.78  &  6.41  &  1.56  &  1.41  &  0.50  &  0.91  &  0.23  &  0.71  \\
1B$_{u}$  &  -616.388094  &  9.39  &  4.32  &  1.44  &  0.80  &  0.52  &  0.48  &  0.24  &  0.34  \\
2B$_{u}$  &  -616.362242  &  8.28  &  3.01  &  1.41  &  0.52  &  0.53  &  0.30  &  0.24  &  0.21  \\
3B$_{u}$  &  -616.357091  &  9.16  &  1.79  &  2.11  &  0.34  &  0.72  &  0.16  &  0.34  &  0.11  \\
3A$_{g}$  &  -616.352480  & 19.75  & 15.46  &  2.83  & 13.78  &  0.79  & 13.49  &  0.38  & 11.57  \\
4A$_{g}$  &  -616.339250  &  7.54  & 20.53  &  1.64  &  3.85  &  0.66  &  0.45  &  0.30  &  0.30  \\
4B$_{u}$  &  -616.326394  & 13.34  & 21.11  &  2.69  &  6.16  &  0.84  &  5.95  &  0.41  &  5.93  \\

\end{tabular}
\end{ruledtabular}
\end{table*}

\begin{figure*}[tb]
\includegraphics[scale=0.25, bb=0 0 1943 608]{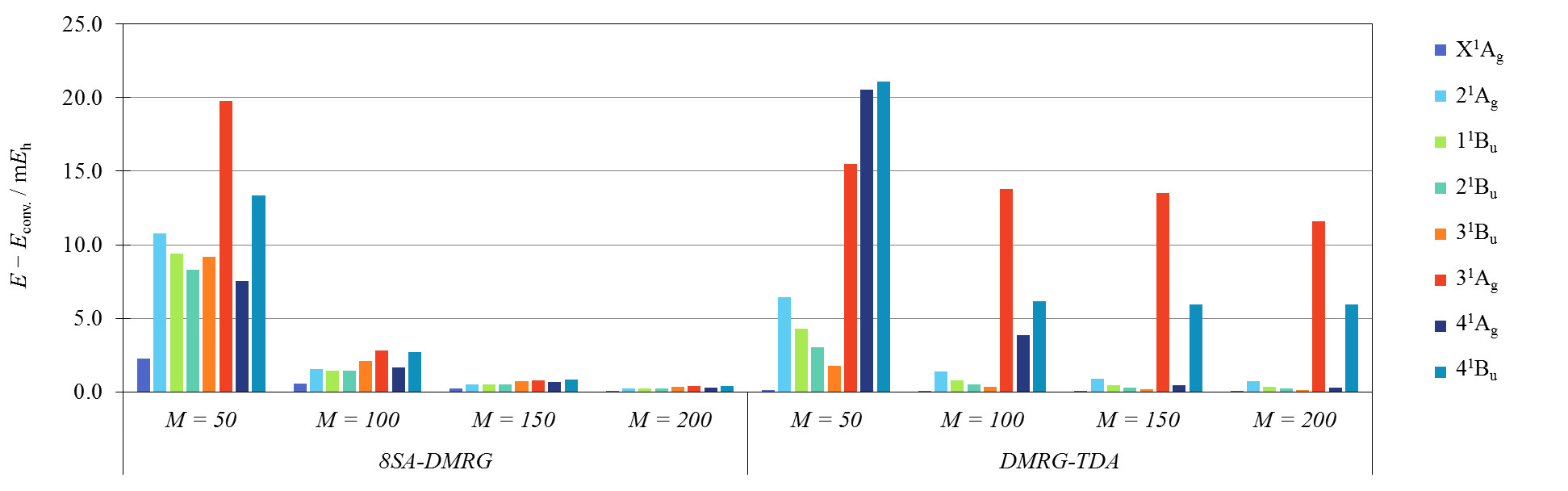}
\caption{Energy errors in m$E_{h}$ for the C$_{16}$H$_{18}$ molecule; a graphical summary for Table \ref{table:m_sweep16}.}
%\caption{Energy errors in m$E_{h}$ for the C$_{16}$H$_{18}$ molecule, computed by 8SA-DMRG and DMRG-TDA with $M$ = 50, 100, 150, and 200. The converged energy $E_{\text{conv.}}$ is computed by an 8SA-DMRG calculation with $M$ = 1000.}
\label{fig:m_sweep16}
\end{figure*}

\section{Illustrative Applications}
In this section, we present ab initio DMRG-TDA benchmark calculations on molecular systems, and compare them to SA-DMRG and other conventional methods. Since we previously reported DMRG-RPA benchmark calculations for the Hubbard and Pariser-Parr-Pople (PPP) model Hamiltonians,\cite{wouters2013} here we only focus on DMRG-TDA.

First, we compute the low-lying excited states of polyenes. DMRG works very well for this type of molecule. Single-reference theories, in contrast, fail due to the doubly excited configurations which appear both in the ground and excited states.

Second, we consider the excited states of the water molecule to investigate the performance of DMRG-TDA for more general non-one-dimensional molecules.

Finally, we test the performance of DMRG-TDA for a highly complex system, the [2Fe-2S] iron-sulfur cluster. It is very difficult to accurately investigate this cluster with conventional theories, due to the near-degeneracy of ground and excited states.

%%
%% Low-Lying Excited States of Polyene Molecules
%%

\subsection{Low-Lying Excited States of Polyenes}

Low-lying excited states of polyenes are of interest because the doubly excited A$_{g}$ state appears to cross the first B$_{u}$ state in longer polyenes. With increasing chain length, the energy gap between the ground state and these excited states becomes smaller, and doubly excited configurations start to mix into the ground state. Non-dynamical electron correlation is therefore important in long polyene molecules. It is well established that DMRG performs well for such a non-dynamical electron correlation, and hence polyene molecules.
% despite the absence of a clear local interaction picture on a one-dimensional lattice. Polyenes are therefore ideal testcases for post-DMRG methods such as DMRG-TDA.

We carried out DMRG-TDA and SA-DMRG calculations for the lowest 8 excited states of C$_{n}$H$_{n+2}$, where $n$ = 4, 8, 12, 16, 20, and 24. Geometries were taken to be all-trans and were optimized at the B3LYP/cc-pVDZ level of theory. Molecular orbitals were computed at the RHF/cc-pVDZ level of theory, and localized for occupied and unoccupied spaces separately. The active space was chosen to be $\pi$-double valence, i.e. $n$ $\pi$-electrons in $2n$ $\pi$-orbitals, where $n$ is the number of carbon atoms.

%% M-sweep for C08H10 and C16H18
To investigate the convergence of energies with $M$, we performed DMRG-TDA and 8SA-DMRG calculations on C$_{8}$H$_{10}$ and C$_{16}$H$_{18}$ with $M$ = 50, 100, 150, and 200. The energy errors from the converged calculations are summarized in Tables \ref{table:m_sweep08} and \ref{table:m_sweep16} (see also Figures \ref{fig:m_sweep08} and \ref{fig:m_sweep16}). From these results we can infer that DMRG-TDA gives faster convergence in the small $M$ region (up to $M = 100$), but slower convergence in the large $M$ region, as compared to 8SA-DMRG. DMRG-TDA works better in the region where $M$ is not sufficiently large to describe the ground and excited states simultaneously with SA-DMRG. In contrast, SA-DMRG can describe higher order excitations from the ground state for which DMRG-TDA converges slow with $M$.
%in the large $M$ region consequently means that higher order corrections must be taken into account to accurately describe these excited states.
An example of this is the 3A$_{g}$ state of C$_{16}$H$_{18}$.
%is converged much slower with DMRG-TDA than with 8SA-DMRG. This excitation is hence delocalized in the DMRG chain over two or more sites.
We conclude that this state involves multi-site excitations in the DMRG chain.

%%
%% Table: 1-site excitation?
%%
\begin{table}[tb]
\caption{Square norm of one-particle transition density matrix $\sum_{pq} \gamma_{pq}^{2}$
         compared to the errors in the DMRG-TDA calculations for C$_{16}$H$_{18}$ ($\Delta E$ in m$E_{h}$) with $M$ = 200.}
\label{table:site_excitation}
\begin{ruledtabular}
\begin{tabular}{ c c c }

state     & $\sum_{pq} \gamma_{pq}^{2}$ & $\Delta E$ \\
\hline
2A$_{g}$  &  0.507 &  0.71 \\
1B$_{u}$  &  0.533 &  0.34 \\
2B$_{u}$  &  0.566 &  0.21 \\
3B$_{u}$  &  1.570 &  0.11 \\
3A$_{g}$  &  0.266 & 11.57 \\
4A$_{g}$  &  0.611 &  0.30 \\
4B$_{u}$  &  0.246 &  5.93 \\

\end{tabular}
\end{ruledtabular}
\end{table}

%It is hard to understand a site-based excitation with standard chemical intuition, because it in general involves a combination of multi-particle excitations. To prove this, we extracted the one-particle character for each state, and compared it with the error in DMRG-TDA. The one-particle transition density matrix
To help analyze the nature of the excited states, we have computed the one-particle transition density matrix
\begin{equation}
  \gamma_{pq}^{\nu} = \braket{\Psi^{\nu}|\hat{a}_{p}^{\dag}\hat{a}_{q}|\Psi_{0}}
\end{equation}
%was computed from the converged SA-DMRG calculation,
and its square norm $\sum_{pq} \left( \gamma_{pq}^{\nu} \right)^{2}$ is summarized in Table \ref{table:site_excitation}.
These square norms show that the 3B$_{u}$ state consists of a single-particle excitation, while the 3A$_{g}$ and 4B$_{u}$ states have a large multi-particle excitation character. Some correlation between the square norms and the errors of DMRG-TDA can be deduced for the polyenes: if the single-particle character of the excitation is large, a single-site excitation (DMRG-TDA) describes it well, and vice versa. Thus in the polyenes, DMRG-TDA works well unless the single-particle character is completely lost.
%To conclude: a single-site excitation captures most of the effect of single-particle excitations, but only part of the multi-particle excitations.
This is consistent with the physics of polyenes: the single excited states, due to the poorly screened Coulomb interaction, consist of strongly bound charged quasiparticles (and are thus linear combinations of ``local'' site excitations) while the doubly excited states consist of weakly bound neutral (triplet excitation) quasiparticles\cite{tavan1987} (and thus involve two independent ``local'' excitations, which cannot be captured well in a single-site picture).

%%
%% Figure: N-sweep
%%
\begin{figure}[tb]
\includegraphics[scale=0.2, bb=0 0 960 720]{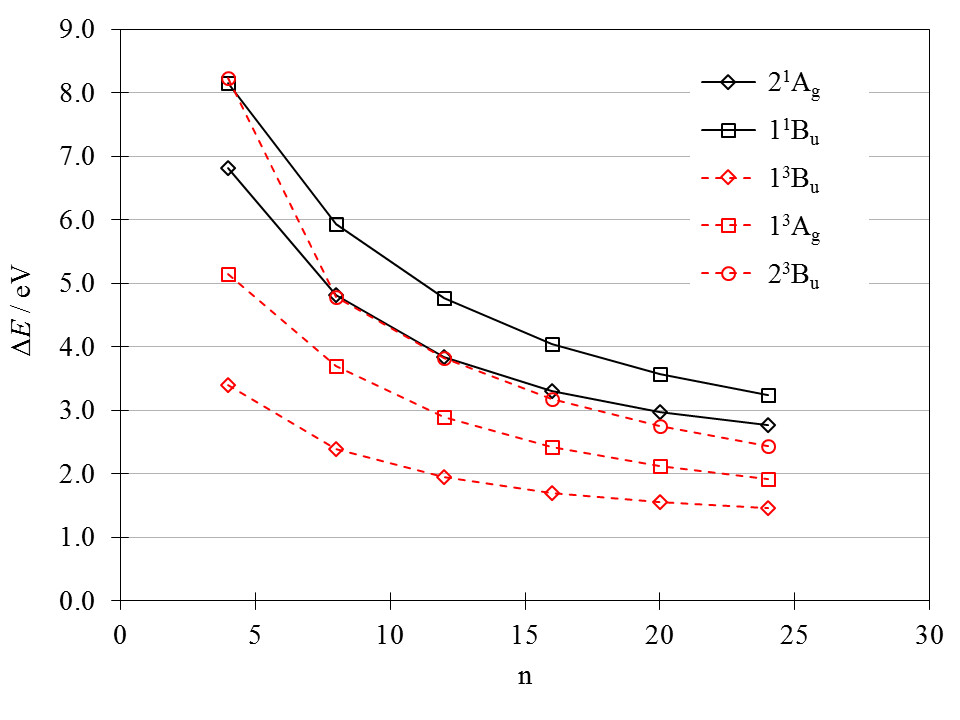}
\caption{Changes of excitation energy in eV for the 2$^{1}A_{g}$, 1$^{1}B_{u}$, 1$^{3}B_{u}$, 1$^{3}A_{g}$, and 2$^{3}B_{u}$ states of polyene molecules C$_{n}$H$_{n+2}$ where $n$ ranges from 4 to 24.}
\label{fig:polyene_n_sweep}
\end{figure}

%% N-sweep for polyene chain
We also investigated the change of the excitation energies with elongation of the polyene chain. Fig.~\ref{fig:polyene_n_sweep} shows how the energy of the lowest 2 singlet excited states and the lowest 3 triplet excited states changes, as computed by DMRG-TDA with $M$ = 200. The curves agree with previous studies,\cite{tavan1987,ohmine1978,ghosh2008} and it can therefore be concluded that DMRG-TDA correctly describes the low-lying excited states of polyene molecules.

%%
%% Water Molecule
%%

\subsection{Water Molecule}

%% TODO: what's the relevant citation for benchmark calcs. of water molecule?
The water molecule is often used as a benchmark system for excited state theories. Here, we present the lowest 12 excited states computed by DMRG-TDA, SA-DMRG, and EOM-CCSD with the cc-pVDZ and aug-cc-pVDZ basis sets.
%The geometry was taken from experiment.\cite{XXX}
The calculations were carried out with the equilibrium geometry; R(OH) = 1.8111$a_{\text{Bohr}}$ and $\angle$HOH = 104.45$^{\circ}$.
The active spaces we employed were 8 electrons in 23 orbitals and 8 electrons in 40 orbitals, for the cc-pVDZ and aug-cc-pVDZ basis sets, respectively. The 1s core orbitals were kept frozen. We performed the DMRG-TDA and 4SA-DMRG calculations with $M$ = 500, for each irreducible symmetry. It should be noted that the reference states were computed separately for each state symmetry in the DMRG-TDA calculation,
because a symmetry changing deviation cannot always be captured by ``single-site'' excitation, due to missing quantum numbers in the ground state as discussed earlier (also see below).

%%
%% Table: H2O / cc-pVDZ
%%
\begin{table}[tb]
\caption{Energies of the ground and the lowest 12 excited states of the water molecule as computed by 4SA-DMRG ($M$ = 500), DMRG-TDA ($M$ = 500), and EOM-CCSD, in the cc-pVDZ basis set. Note that the DMRG-TDA and SA-DMRG calculations were carried out for each irreducible representation separately.}
\label{table:water_vdz}
\begin{ruledtabular}
\begin{tabular}{ c c c c c }

\multirow{3}{*}{state} & $E_{\text{conv.}}$ / $E_{h} $ & \multicolumn{3}{c}{$E - E_{\text{conv.}}$ / m$E_{h}$} \\
\cline{2-2}\cline{3-5}
          & 4SA          & 4SA        &      TDA   & \multirow{2}{*}{EOM-CCSD} \\
          & $M$ = 2000   & $M$ =  500 & $M$ =  500 &  \\
\hline
XA$_{1}$  &  -76.241697  &  0.11  &  0.01  &  3.68 \\
1B$_{1}$  &  -75.939176  &  0.20  &  0.02  &  1.50 \\
1A$_{2}$  &  -75.864445  &  0.20  &  0.02  &  2.07 \\
2A$_{1}$  &  -75.842487  &  0.18  &  0.12  &  2.08 \\
1B$_{2}$  &  -75.765964  &  0.21  &  0.02  &  2.55 \\
2B$_{2}$  &  -75.696018  &  0.20  &  0.45  &  4.14 \\
3A$_{1}$  &  -75.584080  &  0.16  &  0.10  &  5.13 \\
4A$_{1}$  &  -75.462977  &  0.26  &  2.69  &   N/A \\
2A$_{2}$  &  -75.448180  &  0.23  &  0.04  &  4.53 \\
3A$_{2}$  &  -75.403286  &  0.33  &  0.27  &   N/A \\
2B$_{1}$  &  -75.401515  &  0.28  &  0.25  & 24.20 \\
3B$_{1}$  &  -75.381977  &  0.26  &  0.07  &   N/A \\
3B$_{2}$  &  -75.322655  &  0.26  &  0.07  &  5.75 \\

\end{tabular}
\end{ruledtabular}
\end{table}

%%
%% Table: H2O / aug-cc-pVDZ
%%
\begin{table}[tb]
\caption{Energies of the ground and the lowest 12 excited states of the water molecule as computed by 4SA-DMRG ($M$ = 500), DMRG-TDA ($M$ = 500), and EOM-CCSD, in the aug-cc-pVDZ basis set. Note that the DMRG-TDA and SA-DMRG calculations were carried out for each irreducible representation separately.}
\label{table:water_avdz}
\begin{ruledtabular}
\begin{tabular}{ c c c c c }

\multirow{3}{*}{state} & $E_{\text{conv.}}$ / $E_{h} $ & \multicolumn{3}{c}{$E - E_{\text{conv.}}$ / m$E_{h}$} \\
\cline{2-2}\cline{3-5}
          & 4SA          & 4SA        &      TDA   & \multirow{2}{*}{EOM-CCSD} \\
          & $M$ = 2000   & $M$ =  500 & $M$ =  500 &  \\
\hline
XA$_{1}$  &  -76.274423  &  1.29  &  0.31  &  5.86 \\
1B$_{1}$  &  -75.997383  &  2.04  &  0.49  &  2.66 \\
1A$_{2}$  &  -75.931824  &  1.91  &  0.43  &  1.96 \\
2A$_{1}$  &  -75.909074  &  2.17  & 11.51  &  2.84 \\
2B$_{1}$  &  -75.863101  &  2.02  &  1.47  &  2.10 \\
1B$_{2}$  &  -75.844352  &  2.06  &  0.42  &  1.97 \\
3A$_{1}$  &  -75.839232  &  2.46  &  4.47  &  2.63 \\
3B$_{1}$  &  -75.833279  &  2.13  &  0.54  &  2.30 \\
2A$_{2}$  &  -75.826508  &  2.02  &  0.54  &  3.41 \\
3A$_{2}$  &  -75.788484  &  2.02  &  0.49  &  1.29 \\
4B$_{1}$  &  -75.770624  &  3.14  & 11.36  &  2.14 \\
4A$_{1}$  &  -75.766827  &  2.27  &  9.44  &  2.37 \\
2B$_{2}$  &  -75.762108  &  2.00  & 17.53  &  3.89 \\

\end{tabular}
\end{ruledtabular}
\end{table}

%%
%% the DMRG-TDA numbers with LMO
%%
\begin{table}[tb]
\caption{Energies of the ground and the lowest 12 excited states of the water molecule as computed by 8SA-DMRG ($M$ = 500), 13SA-DMRG ($M$ = 500), and DMRG-TDA ($M$ = 500) in the cc-pVDZ basis set, where the valence molecular orbitals were localized. The 4SA $M$ = 2000 results were taken from Table \ref{table:water_vdz}.}
\label{table:water_vdz_lmo}
\begin{ruledtabular}
\begin{tabular}{ c c c c c }

\multirow{3}{*}{state} & $E_{\text{conv.}}$ / $E_{h} $ & \multicolumn{3}{c}{$E - E_{\text{conv.}}$ / m$E_{h}$} \\
\cline{2-2}\cline{3-5}
          & 4SA          & 8SA       & 13SA      & TDA \\
          & $M$ = 2000   & $M$ = 500 & $M$ = 500 & $M$ = 500 \\
\hline
XA$_{1}$  &  -76.241697  &  1.62  &  3.01  &  0.24  \\
1B$_{1}$  &  -75.939176  &  2.11  &  3.37  &  1.20  \\
1A$_{2}$  &  -75.864445  &  2.71  &  4.05  &  2.78  \\
2A$_{1}$  &  -75.842487  &  1.68  &  3.11  &  0.43  \\
1B$_{2}$  &  -75.765964  &  1.96  &  3.64  &  0.57  \\
2B$_{2}$  &  -75.696018  &  1.47  &  2.90  &  0.29  \\
3A$_{1}$  &  -75.584080  &  2.10  &  4.13  &  0.39  \\
4A$_{1}$  &  -75.462977  &  2.86  &  4.77  &  4.89  \\
2A$_{2}$  &  -75.448180  &   --   &  6.78  &  N/A   \\
3A$_{2}$  &  -75.403286  &   --   &  5.50  &  5.39  \\
2B$_{1}$  &  -75.401515  &   --   &  4.31  &  5.37  \\
3B$_{1}$  &  -75.381977  &   --   &  7.90  &  N/A   \\
3B$_{2}$  &  -75.322655  &   --   &  7.00  &  6.98  \\

\end{tabular}
\end{ruledtabular}
\end{table}

Energies of the ground and lowest 12 excited states are summarized in Tables \ref{table:water_vdz} and \ref{table:water_avdz}. With the cc-pVDZ basis set, both the DMRG-TDA and 4SA-DMRG energies are converged at $M$ = 500, except for the 4A$_{1}$ state with DMRG-TDA.
With the larger aug-cc-pVDZ basis set, the 4SA-DMRG numbers exhibit systematic errors on the order of 2 m$E_{h}$, while the DMRG-TDA numbers are generally better than the 4SA-DMRG numbers, except for a few states.
%The errors for DMRG-TDA are however not systematic, indicating that DMRG-TDA breaks down for some specific cases (the 4B$_{1}$, 4A$_{1}$, and 2B$_{2}$ states). These states cannot be described in terms of a ``single-site'' excitation.
The errors for DMRG-TDA are however not systematic, indicating that DMRG-TDA breaks down for some specific cases, e.g. 2A$_{1}$ and 3A$_{1}$ states. These states cannot be described in terms of a ``single-site'' excitation.
%The performance of DMRG-TDA highly depends on the character of the state, and the character of the molecule.

The EOM-CCSD energies mostly agree with the converged energies on the order of 3 m$E_{h}$, but some higher energy states are missing with the cc-pVDZ basis set.
%Because this basis set does not contain any diffuse functions, it is unable to describe Rydberg excitations. It is expected that the unphysical description of such states is far from a "single-particle" excitation. On the other hand,
These high energy states, which have high particle-rank character, are nonetheless correctly described by DMRG-TDA. Conversely EOM-CCSD works very well for certain singly-excited states for which DMRG-TDA breaks down.
This illustrates the fact that single-site and single-particle excitations are in general of fundamentally different character.
%Consequently, the "single-site" excitation in DMRG-TDA can potentially take into account the "multi-particle" excitation, and vice versa. [Remark from Seb: this contradicts the conclusion made for the polyenes]

%We also present DMRG-TDA results when the occupied orbitals are localized, as summarized in Table \ref{table:water_vdz_lmo}. In this calculation, the A$_{1}$ ground state was chosen as the reference for all excited states, i.e. symmetry changing excitations were considered. DMRG-TDA works better if localized orbitals are used instead of canonical orbitals, for most of the low-lying excited states of water molecule. However, there are some missing excited states that appeared in the SA-DMRG calculation. This indicates that symmetry changing excitations cannot always be possible to be represented by single-site excitations because the ground-state may be missing quantum numbers important to the symmetry change.
We also present DMRG-TDA results when the valence orbitals are localized, as summarized in Table \ref{table:water_vdz_lmo}. In this calculation, the A$_{1}$ ground state was chosen as the reference for all excited states, i.e. symmetry changing excitations were considered. DMRG-TDA works better than SA-DMRG if localized orbitals are used instead of canonical orbitals, for most of the low-lying excited states of the water molecule. However, there are larger errors in high energy states and some missing excited states. This indicates that symmetry changing excitations cannot always be represented by single-site excitations because the ground-state may be missing quantum numbers important to the symmetry change.

%%
%% Iron-Sulfur Cluster
%%

\subsection{[Fe$_{2}$S$_{2}$(SCH$_{3}$)$_{4}$]$^{3-}$ Cluster}

%%
%% Table: [Fe2S2] Cluster
%%
\begin{table*}[tb]
\caption{Excitation energies in eV of the lowest 10 states for the doublet, quartet, sextet, and octet spin states, computed by 10SA-DMRG and DMRG-TDA with $M$ = 500. The active space involves 3d orbitals of the iron centers and 3p orbitals of the sulfur atoms, except for the non-bonding orbitals. In total, it consists of 31 electrons in 20 orbitals. The ground state is in the doublet spin state.}
\label{table:fe2s2_svp}
\begin{ruledtabular}
\begin{tabular}{ c c c c c c c c c }

\multirow{2}{*}{state} & \multicolumn{4}{c}{10SA-DMRG ($M$ = 500)} & \multicolumn{4}{c}{DMRG-TDA ($M$ = 500)} \\
\cline{2-5}\cline{6-9}
     & doublet& quartet& sextet & octet  & doublet& quartet& sextet & octet  \\
\hline
 1A  &  0.00  &  0.02  &  0.06  &  0.09  &  0.00  &  0.02  &  0.05  &  0.10  \\
 2A  &  0.04  &  0.06  &  0.08  &  0.13  &  0.03  &  0.06  &  0.07  &  0.14  \\
 3A  &  0.14  &  0.17  &  0.28  &  0.44  &  0.14  &  0.16  &  0.28  &  0.45  \\
 4A  &  0.31  &  0.46  &  0.52  &  0.59  &  0.31  &  0.46  &  0.52  &  0.59  \\
 5A  &  0.45  &  0.48  &  0.53  &  0.62  &  0.45  &  0.48  &  0.53  &  0.63  \\
 6A  &  0.52  &  0.51  &  0.57  &  0.68  &  0.52  &  0.51  &  0.56  &  0.69  \\
 7A  &  0.54  &  0.52  &  0.68  &  0.95  &  0.54  &  0.51  &  0.68  &  0.96  \\
 8A  &  0.54  &  0.66  &  0.82  &  1.02  &  0.54  &  0.65  &  0.81  &  1.04  \\
 9A  &  0.67  &  0.80  &  0.96  &  1.03  &  0.68  &  0.80  &  0.96  &  1.04  \\
10A  &  0.71  &  0.85  &  1.03  &  1.11  &  0.72  &  0.86  &  1.03  &  1.17  \\

\end{tabular}
\end{ruledtabular}
\end{table*}

\begin{figure}[tb]
\includegraphics[scale=0.25, bb=0 0 720 680]{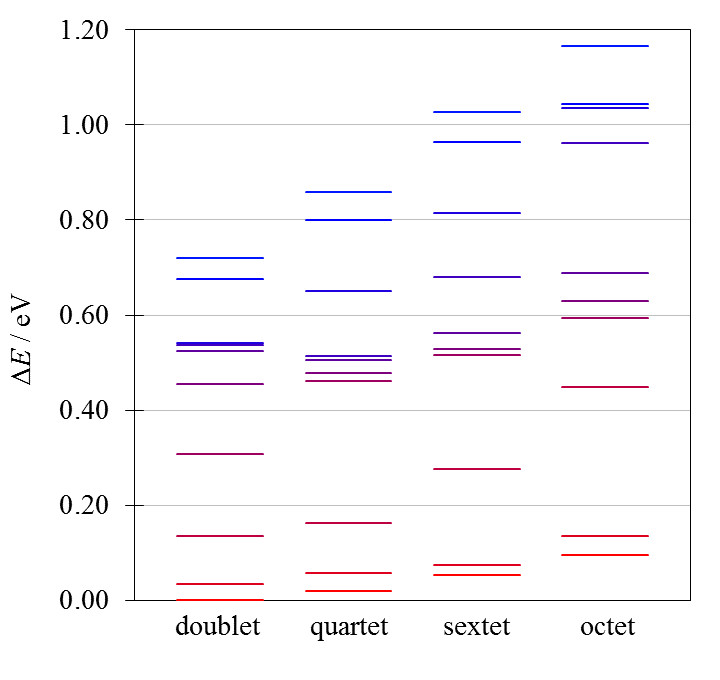}
\caption{Excitation energies in eV of the lowest 10 states for the doublet, quartet, sextet, and octet spin states, computed by DMRG-TDA with $M$ = 500. See also Table~\ref{table:fe2s2_svp} for more detailed data.}
\label{fig:fe2s2ex}
\end{figure}

The [2Fe-2S] iron-sulfur clusters are found in various classes of oxidoreductase enzymes, which mediate electron transfer from a redox molecule such as NAD$^{+}$/NADH to the enzyme reaction center.

Recently, a high-accuracy DMRG calculation on such an iron-sulfur cluster was performed and helped to clarify the ground state electronic structure of the complicated cluster.\cite{sharma2012} The excited states of the iron-sulfur cluster are also interesting, since the ground and excited states are expected to be highly degenerate. Here, we demonstrate the performance of SA-DMRG and DMRG-TDA for the large number of quasi-degenerate excited states.

Geometry, basis sets, and active space were taken from the earlier DMRG work.\cite{sharma2012} We focused on the doublet, quartet, sextet, and octet spin states and computed the lowest 10 states for each spin state.

Table~\ref{table:fe2s2_svp} summarizes the excitation energies of the [Fe$_{2}$S$_{2}$(SCH$_{3}$)$_{4}$]$^{3-}$ cluster. The SA-DMRG and DMRG-TDA energies are very close to each other, which implies that for this molecule SA-DMRG surprisingly works as well as DMRG-TDA, despite the averaging over many states. It indicates that the renormalized basis necessary for a good ground-state description, is also relevant for the excited states.

The excited states of [Fe$_{2}$S$_{2}$(SCH$_{3}$)$_{4}$]$^{3-}$ are very close to the ground state, and conventional single reference theories have great difficulty in describing such a system due to the serious quasi-degeneracy (see Fig.~\ref{fig:fe2s2ex}). It is interesting to see that only ``single-site'' excitations from the DMRG reference wavefunction are sufficient to compute these complicated excited states.

%%
%% Conclusion
%%

\section{Summary}

In this work, we discussed in detail two post-density matrix renormalization group (DMRG),
namely, the DMRG analogue of the Tamm-Dancoff Approximation (DMRG-TDA) and the DMRG analogue of the Random Phase Approximation (DMRG-RPA),
which were introduced both in our earlier work (Ref.~\onlinecite{wouters2013}) and the work of Haegeman et al (Ref.~\onlinecite{haegeman2013}).
These methods provide new routes to excited states within DMRG.
Both can be derived from our earlier linear response theory for the DMRG (Ref.~\onlinecite{dorando2009}).
One of the main purposes of this work was to present an efficient sweep algorithm to solve for excited states in the DMRG-TDA and DMRG-RPA equations.
The algorithm we presented  may be easily implemented within any existing DMRG code, thus opening up the simple
adoption of these techniques.

We further presented benchmark calculations on a number of ab initio model systems: polyenes, the water molecules, 
and a [2Fe-2S] cluster.
These calculations provide insight into the ``single-site'' nature of excitations in DMRG-TDA;
single-site meaning that the excitation is generated by a sum of operators, each acting on a single site site or small number of consecutive sites in the DMRG lattice.
In particular, single-{\it site} excitations do not generally correspond
to single-{\it particle} excitations in particle-based theories. Rather, some many-particle excitations
are easily described with single-site excitations, while other single-particle excitations are hard to describe.
DMRG-TDA (and DMRG-RPA) thus offer a complementary approach to excited states, as compared to standard particle-based theories.
Whether or not an excited state is well described by DMRG-TDA 
also provides useful physical insight. For example, in the case of polyenes, DMRG-TDA provided a good description
of singly-excited states and a poor description of doubly-excited states, suggesting that
singly-excited states consist of strongly bound quasi-particles, while
the doubly-excited states consist of weakly bound quasi-particles. 

The DMRG-TDA and DMRG-RPA methods discussed  here are the lowest rung on a more general hierarchy
of post-DMRG methods discussed in Ref.~\onlinecite{wouters2013}. Their success and complementary nature 
to standard particle-based mean-field hierarchies provides further motivation to explore higher rungs on the 
post-DMRG ladder.

\section{Acknowledgements}
N.N. would like to thank Dr. Sandeep Sharma for his help on the computations of the iron-sulfur cluster. S.W. acknowledges funding from the Research Foundation Flanders.
This work was supported by the National Science Foundation (NSF) through Grant No. NSF:SSI-SSE, OCI-1265278 and NSF:CHE-1265277.

%%
%% Bibliography
%%

%\bibliographystyle{apsrev4-1}
\bibliography{lrt_ref}

\end{document}